\documentclass[runningheads]{llncs}
\usepackage[T1]{fontenc}

\usepackage{color}
\usepackage{xcolor, colortbl}
\usepackage{makecell, booktabs, caption}
\usepackage[noend]{algpseudocode}
\usepackage{textcomp}
\usepackage{listings}
\usepackage{hyperref}
\usepackage{alltt}
\usepackage{tikz}
\usepackage{framed}
\usepackage{mdframed}
\usepackage{marvosym}
\usepackage[nointegrals]{wasysym}
\usepackage{marvosym}
\usepackage{crayola}
\usepackage{mathpartir}
\usepackage{tabularx}
\usepackage[belowskip=-15pt,aboveskip=0pt]{caption}
\usepackage[skins]{tcolorbox}
\usepackage{multicol}
\usepackage{import}
\usepackage[ruled,vlined]{algorithm2e}
\usepackage{algpseudocode}
\usepackage{amsmath}
\usepackage{lipsum}
\usepackage{amsfonts}
\usepackage{amssymb}
\usepackage{cite}	
\usepackage{svg}
\usepackage{pdflscape}
\usepackage{rotating}

\usetikzlibrary{positioning,shapes,arrows, backgrounds, fit, shadows}
\usetikzlibrary{arrows.meta}
\usetikzlibrary{decorations.markings}

\algnewcommand\algorithmicforeach{\textbf{for each}}
\algdef{S}[FOR]{ForEach}[1]{\algorithmicforeach\ #1\ \algorithmicdo}

\SetCommentSty{mycommfont}

\SetKwInput{KwInput}{Input}                
\SetKwInput{KwOutput}{Output}              

\lstdefinestyle{javastyle}{
    language = java,
	basicstyle = \ttfamily\scriptsize,
	stringstyle = \ttfamily,
	keywordstyle=\color{Blue}\bfseries,
	identifierstyle=\color{mauve},
	commentstyle=\color{OliveGreen},
	frameround=tttt,
	showstringspaces=false,
	captionpos=b
}

\lstdefinestyle{code}{frame=tb,
  language=Java,
  aboveskip=3mm,
  belowskip=3mm,
  showstringspaces=false,
  columns=flexible,
  basicstyle={\small\ttfamily},
  numbers=none,
  numberstyle=\tiny\color{gray},
  keywordstyle=\color{blue},
  commentstyle=\color{dkgreen},
  stringstyle=\color{mauve},
  breaklines=true,
  breakatwhitespace=true,
  tabsize=3
}

\tikzstyle{bb}=[%
      rectangle, draw=black, thick, fill=OliveGreen!30, drop shadow, align=center,
      text ragged, minimum height=2em, minimum width=2em, inner sep=6pt
]

\tikzstyle{inv}=[%
      rectangle, draw=none,  align=center,
      text ragged, minimum height=2em, minimum width=2em, inner sep=6pt
]

\tikzstyle{db}=[%
      ellipse, draw=black, thick, fill=pink, drop shadow, align=center,
      text ragged, minimum height=2em, inner sep=6pt
]

\tikzstyle{jn}=[%
      ellipse, draw=black, thick, fill=black
]

\tikzstyle{io}=[%
      trapezium, trapezium left angle=60, trapezium right angle=120, draw=black, thick, fill=brown, drop shadow,
      text ragged, minimum height=2em, minimum width=2em, inner sep=6pt, align=center
]

\tikzstyle{glio}=[%
      trapezium, trapezium left angle=60, trapezium right angle=120, draw=red, line width = 1mm, fill=brown, drop shadow,
      text ragged, minimum height=2em, minimum width=2em, inner sep=6pt
]
\tikzstyle{gl}=[%
      rectangle, draw=red, line width = 1mm, fill=lightblue, drop shadow,
      text ragged, minimum height=2em, minimum width=2em, inner sep=6pt
]

\tikzstyle{en}=[%
      rectangle, draw=black, thick, fill=none,
      text ragged, minimum height=2em, minimum width=2em, inner sep=6pt
]

\tikzstyle{sh}=[%
      rectangle, draw=gray, thick, fill=none, color = gray,
      text ragged, minimum height=2em, minimum width=2em, inner sep=6pt
]

\tikzset{obj/.style = {rectangle split, rounded corners,
                      rectangle split parts=2, very thick,draw=black!50, top
                      color=white,bottom color=black!20, align=center}}

\tikzset{cobj/.style = {rectangle split,
                      rectangle split parts=3, very thick,draw=black!50, top
                      color=white,bottom color=black!10, align=left}}

\newtcolorbox{myframe}[2][]{enhanced,colback=white,colframe=black,coltitle=black, sharp corners,boxrule=0.4pt,
  fonttitle=\itshape, attach boxed title to top left={yshift=-0.3\baselineskip-0.4pt,xshift=2mm}, boxed title style={tile,size=minimal,left=0.5mm,right=0.5mm, colback=white,before upper=\strut}, title=#2,#1}

\definecolor{dkgreen}{rgb}{0,0.6,0}
\definecolor{gray}{rgb}{0.5,0.5,0.5}
\definecolor{mauve}{rgb}{0.58,0,0.82}

\begin{document}
\setlength{\abovedisplayskip}{0pt}
\setlength{\belowdisplayskip}{0pt}
\setlength{\abovedisplayshortskip}{0pt}
\setlength{\belowdisplayshortskip}{0pt}
\title{ConStaBL - A Fresh Look at Software Engineering with State Machines}
\author{Karthika Venkatesan\inst{1,2} \and
Sujit Kumar Chakrabarti\inst{1}}
\authorrunning{Karthika et al.}
\institute{International Institute of Information Technology Bangalore, INDIA \email{\{karthika.venkatesan,sujitkc\}@iiitb.ac.in}
\and
Centre for Development of Advanced Computing Bangalore, INDIA}
\maketitle

\begin{abstract}
Statechart is a visual modelling language for systems. In this paper, we extend our earlier work on modular statecharts with local variables and present an updated operational semantics for statecharts with concurrency. Our variant of the statechart has local variables, which interact significantly with the remainder of the language semantics. Our semantics does not allow transition conflicts in simulations and is stricter than most other available semantics of statecharts in that sense. It allows arbitrary interleaving of concurrently executing action code, which allows more precise modelling of systems and upstream analysis of the same. We present the operational semantics in the form of the simulation algorithm. We also establish the criteria based on our semantics for defining conflicting transitions and valid simulations. Our semantics is executable and can be used to simulate statechart models and verify their correctness. We present a preliminary setup to carry out fuzz testing of Statechart models, an idea that does not seem to have a precedent in literature. We have used our simulator in conjunction with a well-known fuzzer to do fuzz testing of statechart models of non-trivial sizes and have found issues in them that would have been hard to find through inspection.
\end{abstract}
\section{Introduction}
\label{introduction}
Statecharts have been a popular modelling notation for several decades now. Many implementations are in use: Stateflow\cite{stateflow}, Yakindu\cite{yakindu}, Boost\cite{States2006}, Sismic\cite{Decan2020}, Raphsody\cite{10.1007/3-540-47884-1_1}, QM\cite{qm}, Specgen\cite{10.1007/978-3-319-57288-8_19}, and Uppaal\cite{larsen1997uppaal}, both in the commercial and free domains, to name a few. There have been numerous surveys on statecharts so far, indicating their effectiveness and usefulness in modelling complex systems\cite{von1994comparison}\cite{10.1145/3579821}\cite{DBLP:journals/corr/cs-SE-0407038}. Semantics, formal verification, and testing of statecharts are subjects that have been extensively studied on various statechart variants. Although the research on Statechart notation seems to have visibly slowed down over the last decade or so, we think that there are a number of loose ends to tie before we call the work complete.
\begin{enumerate}
\item The inherent complexity of Statechart models needs efforts in the direction of simplification and moduralisation of notation to make Statechart modelling a scalable practice.
\item The semantics of widely available distributions have important shortcomings \cite{6754239}. These need investigation and proposals for mitigation.
\item Due to their higher level of abstraction and complex structure, Statecharts need sophisticated verification, testing, and simulation capabilities to be integrated as a part of the modelling environment.
\end{enumerate}

In this paper, we present ConStaBL (\textbf{Con}current \textbf{Sta}te \textbf{B}ased on \textbf{L}anguage) a statechart variant that includes local variables. Local variables increase modularity but have a bearing on the semantics of the rest of the Statechart language. Hence, we take a fresh look at the operational semantics of ConStaBL. Through this work, we make the following contributions:

\begin{enumerate}
\item ConStaBL: an extension of StaBL, a state-based specification language with local variables, to include concurrency.
\item Operational semantics of ConStaBL.
\item Simulator for ConStaBL.
\item Fuzz testing of Statecharts, presented as an application of the simulator.
\end{enumerate}

We present a novel statechart language that serves as both a high-level modelling and rapid prototyping language. As a high-level modelling language, it enables the construction of abstract representations of the system's states, transitions, and events, taking inspiration from design integrated development tools like Simulink\cite{StateDiagram}, LabView\cite{labview}, and many other statechart modelling tools\cite{10.1145/3579821}. This abstraction enables a clear visualisation of the system's functionality and behaviour to capture complex system dynamics and specify desired system behaviours in a concise and intuitive manner. In addition, our statechart language is an effective rapid prototyping tool as it allows system designers to rapidly run simulations. This feature enables developers to rapidly validate and evaluate the system's behaviour, test various scenarios, and identify potential issues or enhancements during the earliest stages of development. However, there are notable distinctions in our approach to handling the action language, as we have discussed in Section \ref{background}.  

We start by presenting the \textit{language} of ConStaBL statecharts in Section \ref{language} that details the \textit{Abstract Syntax} and \textit{Structural Semantics}. The definitions in this section are described with a common example in Fig. \ref{f:conf-code-example}. The operational semantics are described in Section \ref{o-semantics}. We present the design details of a simulator for this language and an illustration of how our simulator can be used by performing fuzz testing on statecharts using our simulator in Section \ref{fuzzing}.

\section{Related Works}
\label{background}
David Harel introduced statecharts\cite{HAREL1987231} in 1987 as a visual modeling language for complex reactive systems. While there is no consensus on a "one right way" to build semantics for statecharts\cite{harel1996statemate}, various approaches have been proposed to accurately capture system behaviour, given their ability to model real-time, event-driven systems. Over time, more than 100 variants\cite{von1994comparison, 10.1145/3579821, DBLP:journals/corr/cs-SE-0407038, VanMierlo2020} of statecharts have been proposed, each with different features, semantics, and domains.

As research on the execution semantics of statecharts progressed\cite{harel1996statemate}\cite{harel2004rhapsody}\cite{uselton1994process}\cite{1357344}\cite{yakindu}\cite{Eshuis2009}, the topic of concurrency-related semantics became significant. Two major divisions of statechart semantics, namely interleaving and true concurrency\cite{von1994comparison}, have been proposed. Interleaving semantics use priorities to ensure determinism and resolve data races. Transition priorities determine the execution preference in cases of non-determinism, while sequential ordering of substates within an $AND$ composition determines the execution order in the presence of concurrent enabled transitions. This mechanism, though widely used in commercial tools like Yakindu, StateMate, and Stateflow, may limit the analysis of more realistic behaviour in concurrent systems that use multiprocessing architectures. The semantics of UML 2.5.1 mentions that the order in which the enabled transitions in an orthogonal region are executed is left undefined. Whereas, certain variants of statecharts and tools\cite{yakindu} that follow UML semantics, requires the addition of priorities and sequential ordering during design. These design decisions perculates into implementation, as these tools generate sequential code from statecharts (to imperative languages like C, Java, and TypeScript etc.). This may not align with the original intention of designing and analysing concurrent systems. Rhapsody\cite{harel2004rhapsody}, an execution semantics of UML, allows modeling of non-prioritised orthogonal states and acknowledges the undetermined execution order among orthogonal siblings. The arbitrary order of execution is achieved by "locking" each $AND$ state once a transition is triggered within one of its components\cite{ibm_rhapsody}. Raphsody insists that tools should to permit design without transition priorities as it is impossible to determine during compile time, if the guards of two enabled transitions may evaluate to true, though they may respond to the same trigger/ event. SCXML\cite{barnett2017introduction} resolves the execution order in parallel states by its document order. Sismic\cite{Decan2020}, an execution engine for SCXML (that also support majority of UML features) takes a different approach by processing enabled transitions based on the decreasing order of the depth of their source states. This aligns with the inner-first and source-state semantics, processing transitions from deeper states before those from less nested ones. In case of ties, the lexicographic order of the source state names is considered. It also prefers to flag non-determinism rather than following a priority based design.  As far as we are aware, none of these semantic approaches provide specific details on how to interleave the action code.

Our focus is primarily on analysing concurrency behaviour in ConstaBL statecharts with local variables in a priority-agnostic manner. The presence of local variables in our language makes it important to dvelve into the detailed semantics at the action code level. We provide a detailed perspective on how action code within AND compositions is executed in the interleaved manner at the statement level.  While our approach may appear close to tokenized mechanism of Petri nets\cite{peterson1977petri}, there is a significant distinction. Each node in the Code graph that we construct is a Control flow graph on its own. Also, nesting an $AND$ state within another, the composition of action blocks combines sequential and concurrent patterns which needs a special attention. Our methodology formalises this process and enables the early detection and analysis of concurrency-related issues. 

\section{The Language}
\label{language}
In this section, we present the upgrades to the abstract syntax and structural semantics of StaBL\cite{10.1145/3385032.3385040} with the constructs and terminologies that are necessary to discuss the concurrency semantics of \emph{ConStaBL}.
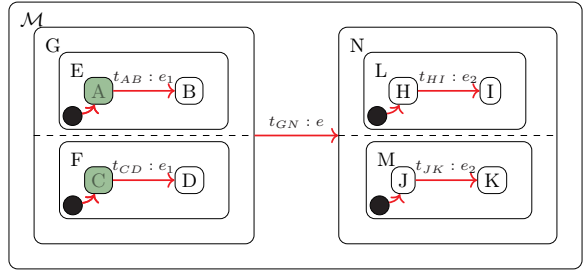
\begin{figure}
    \begin{center}
    \begin{minipage}{0.8\textwidth}
    \centering
    \resizebox{0.8\textwidth}{!}{
    \begin{tikzpicture}
    \node[rectangle, draw, fill=OliveGreen, fill opacity=0.5, rounded corners](a){A};
    \node[rectangle, draw, rounded corners, right=of a](b){B};
    \node[rectangle, draw, fill=OliveGreen, fill opacity=0.5, rounded corners, below=of a](c){C};
    \node[rectangle, draw, rounded corners, right=of c](d){D};
    \node[rectangle, draw, rounded corners, inner sep = 0.4cm, fit=(a) (b)](e){};
    \node[below right=0.1cm of e.north west](le){E};
    \node[circle, draw, fill=Black, minimum size=0.5pt, below left=0.1cm of a](ie){};
    \node[rectangle, draw, rounded corners, inner sep = 0.4cm, fit=(c) (d)](f){};
    \node[below right=0.1cm of f.north west](lf){F};
    \node[circle, draw, fill=Black, minimum size=0.5pt, below left=0.1cm of c](if){};
    
    \node[rectangle, draw, rounded corners, inner sep = 0.4cm, fit=(e) (f)](g){};
    \node[below right=0.1cm of g.north west](lg){G};
    
    \draw[-, dashed] (g.west) -- (g.east);
    
    \draw[->, Red, thick, bend right](ie) to (a);
    \draw[->, Red, thick, bend right](if) to (c);
    
    \node[rectangle, draw, rounded corners, right=3cm of b](h){H};
    \node[rectangle, draw, rounded corners, right=of h](i){I};
    \node[rectangle, draw, rounded corners, below=of h](j){J};
    \node[rectangle, draw, rounded corners, right=of j](k){K};
    \node[rectangle, draw, rounded corners, inner sep = 0.4cm, fit=(h) (i)](l){};
    \node[below right=0.1cm of l.north west](ll){L};
    \node[circle, draw, fill=Black, minimum size=0.5pt, below left=0.1cm of h](il){};
    \node[rectangle, draw, rounded corners, inner sep = 0.4cm, fit=(j) (k)](m){};
    \node[below right=0.1cm of m.north west](lm){M};
    \node[circle, draw, fill=Black, minimum size=0.5pt, below left=0.1cm of j](im){};
    
    \node[rectangle, draw, rounded corners, inner sep = 0.4cm, fit=(l) (m)](n){};
    \node[below right=0.1cm of n.north west](ln){N};
    
    \node[rectangle, draw, rounded corners, inner sep = 0.4cm, fit=(g) (n)](mo){};
    \node[below right=0.1cm of mo.north west](lmo){$\mathcal{M}$};
    
    \draw[-, dashed] (n.west) -- (n.east);
    
    \draw[->, Red, thick, bend right](ie) to (a);
    \draw[->, Red, thick, bend right](if) to (c);
    \draw[->, Red, thick](a) to node[above]{{\scriptsize \color{Black}$t_{AB}: e_1$}}(b);
    \draw[->, Red, thick](c) to node[above]{{\scriptsize \color{Black}$t_{CD}: e_1$}}(d);
    
    \draw[->, Red, thick, bend right](il) to (h);
    \draw[->, Red, thick, bend right](im) to (j);
    \draw[->, Red, thick](h) to node[above]{{\scriptsize \color{Black}$t_{HI}: e_2$}}(i);
    \draw[->, Red, thick](j) to node[above]{{\scriptsize \color{Black}$t_{JK}: e_2$}}(k);
    \draw[->, Red, thick](g) to node[above]{{\scriptsize \color{Black}$t_{GN}: e$}}(n);    
    \end{tikzpicture}
    }
\end{minipage}
\end{center}
\caption{A concurrent statechart model in a configuration $\mathbb{C}=\{A, C\}$}
\label{f:constabl-example}
\end{figure}
\subsection{Abstract Syntax}
A statechart model consists of three components: a set of states (S), a set of transitions (T), and a set of events (E).

A \emph{\textbf{state}} is a \textit{tuple} of $(p,I,\mathcal{V}_l,\mathcal{V}_p,\mathcal{V}_s,a_N,a_X,\tau)$. Here, $\mathcal{V}_l,\mathcal{V}_p,\mathcal{V}_s$ are the variable sets declared within the state with storage classes: parameter, local, and static, respectively. $a_N, a_X$ are the entry and exit actions. $I$ is the set of default or initial substates of the state. $\tau \in \{statechart, atomic, composite, shell\}$ is the type of the state.

A \emph{\textbf{transition}} is a \textit{tuple} of $(p,s,d,e,g,a)$. A transition is annotated as $"e[g]/a"$ on the arrow that connects the \emph{source}(s) and \emph{destination}(d) states. $e$ is the event, $g$ is a guard or condition that has to be $true$, and $a$ is the action code.

In these \textit{tuples}, $p$ is the parent state of the state/ transition and $a$ is an arbitrary piece of code in an imperative programming language, as shown in Listing \ref{lst:action}. 
\begin{lstlisting}[mathescape=true, caption={A minimal representative action language.}, label={lst:action}, basicstyle=\small]
B::={slist}   
slist::= $\epsilon$|s;slist 
s::= v:=e|if(cond) then $B_1$ else $B_2$|while(cond) $B_1$ 
e::= e+e|e*e|e-e|e/e|v|const|-e|function(a1, a2,..)
//slist: statement list, s: statement, v: variable, e: expression
\end{lstlisting}
A statechart model has only one state of the statechart type, which is itself. The states contained within \emph{shell} execute concurrently. A \textit{shell} state bears resemblance to an \textit{AND} state, but it distinguishes itself in terms of how transitions are defined as described in Section \ref{s:s-semantics}. The substates of the \emph{shell} state are of type \emph{composite}, aka. \emph{regions}, and all of them will be active during execution, so $I$ will contain all of its substates. For a composite state, $I$ is one of its substates, and for an atomic state, $I$ is $empty$. Execution semantics are detailed in Section \ref{o-semantics}.
\begin{example}In the $\mathcal{M}$ in Fig. ~\ref{f:constabl-example} is of type \emph{statechart}. The state \emph{$G$} - a \emph{shell} state is represented as $G: (\mathcal{M}, \{E,F\}, \{x_1\},\{p_1\},\{n_1\},\{x_1:=0;p_1:=0;n_1:=0\},\{x_1:=1;p_1:=1;n_1:=1\},shell)$. $I$ corresponds to $\{E,F\}$ (all substates of $G$). $t_{AB}$ is a transition from source state $A$ to destination state $B$, denoted as $t_{AB}: e_1[x_1=1]/{p_2:=1}$. Here, $t_{AB}$ is the name of the transition, $e_1$ is the event, $x_1=1$ is the guard, and ${p_2:=1}$ is the transition action. \qed
\end{example} 
\subsection{Structural Semantics}
\label{s:s-semantics}
\subsubsection{Containment} (denoted by $\prec$) is function between two states $s_1,s_2$, denoted by $s_1 \prec_i s_2$, where $i$ is the level of containment. Containment is antisymmetric (i.e., $\forall s_1, s_2 \in S, (s_1 \prec_* s_2) \land (s_2 \prec_* s_1) \implies (s_1 = s_2)$), not reflexive (i.e., $\forall s \in S, (s \not\prec_* s)$) and transitive (i.e., $\forall s_1, s_2, s_3 \in S, (s_1 \prec_* s_2) \land (s_2 \prec_* s_3) \implies (s_1 \prec_* s_3)$). When $i = 1$, $s_2$ is the \emph{parent} of $s_1$ and $s_1$ is \emph{child/substate} of $s_2$. When $i = * $, $s_2$ is the \emph{ancestor} of $s_1$ and $s_1$ is \emph{descendent} of $s_2$ at an arbitrary level.

Only a few type containments are valid. There are 4 state types: \textit{statechart}, \textit{shell}, \textit{composite} and \textit{atomic}. The root state of a model must be of type \textit{statechart}, and it can have substates of any type but itself. A \textit{composite} state must have substates of type \textit{composite}, \textit{shell}, or \textit{atomic}. A shell state must have substates of type \textit{composite} only. An \textit{atomic} state cannot have substates. \textbf{Substates} is a function that gives the set of all children of state $s$ (i.e.,  $substates(s)=\{s' \in S | s' \prec_1 s\}$).
\subsubsection{Common Ancestors(CA)} For a set of states, $S' = \{s_1,s_2...,s_n\}$, $CA(S')$ is the set of states from $S$ that is an $ancestor$ of all the states in $S'$.
\begin{align*}
&CA(\{s_1,s_2,... s_n\})=\{ s \in S | s_1\prec_* s \land s_2\prec_* s\ \land .... s_n\prec_* s\}
\end{align*}
\subsubsection{Closest common ancestors(CCA)} (denoted by $\sqcup$) of two states $s_1,s_2$ (denoted by $s_1 \sqcup s_2$ \footnote{infix short-form of $\sqcup(\{s_1, s_2\})$}) is a state $s$ from the $\textbf{common ancestors}$ set,which is an ancestor of both $s_1$ and $s_2$ such that it is not the ancestor of any other common ancestor ($s'$). 
\begin{align*}
&s_1 \sqcup s_2=s | s \in CA(\{s_1,s_2\}) \wedge \nexists s' \in CA(\{s_1,s_2\}) \wedge s'\prec_* s \\
&\sqcup(\{s_1,s_2,..,s_n\}) = s | s \in CA(s_1,s_2...s_n) \wedge \nexists s' \in CA(s_1,s_2,.. s_n) \wedge s'\prec_* s
\end{align*}
CCA of set of transitions is the CCA of the source and destination of those transitions (i.e., $CCA(\{t_1,t_2...,t_n\})=\sqcup(\{t_1.s, t_1.d, t_2.s, t_2.d,...,t_n.s, t_n.d\})$).
\subsubsection{Interlevel transitions.} When the parents of the source and destination of a transition are not the same, it is an interlevel transition (i.e. $\exists t \in T | t.s.p \neq t.d.p $).
\begin{enumerate}
 \item There can be no inter-level transitions between the regions of a shell state (i.e., $\nexists t \in T | \sqcup(\{t.s,t.d\}).\tau = shell$). 
 \item There can be no transition between a descendent and an ancestor (i.e., $\nexists t \in T, (t.s \prec_* t.d) \vee (t.d \prec_* t.s)$).  \item The state of type statechart cannot have any incoming or outgoing transitions (i.e., $\nexists t \in T, t.s.\tau \vee t.d.\tau = statechart$).
\end{enumerate}
\begin{example} 
In Fig. ~\ref{f:constabl-example}, $\mathcal{M} \prec_1 G$ means $\mathcal{M}$ is the parent of $G$ and $substates(\mathcal{M}) = \{G,N\}$. Also, $\mathcal{M}$ is the ancestor of states like $E, L, M$ and $F$ (it is also the ancestor of all the states contained within these states). Here, $E.\tau = F.\tau = composite$ and $G.\tau = shell$. 
\begin{align*}
&CA(\{A,D\})=\{G,\mathcal{M}\} \ and \ CA(\{A,J\})=\{\mathcal{M}\}\\
&\sqcup(\{A,B,C,D\}) = G\\
&CCA(\{t_{AB}\}) = \sqcup(\{A,B\}) = E\\
&CCA(\{t_{AB},t_{CD}\}) = \sqcup(\{A,B,C,D\}) = G 
\end{align*} \qed
\end{example}

\section{Operational semantics}
\label{o-semantics}
StaBL statecharts have integrated action language and variable scoping. During execution, processing an event starts by identification of transitions which are \emph{triggered} by an \emph{event} for which the \emph{guard} evaluates to \emph{true}. Subsequently execution proceeds from the source state to the destination state through various intermediate configurations by executing the corresponding action blocks of the states and identified transitions.

\subsubsection{Environment($\sigma$)} The binding of variables to current values is given by $\sigma$. $\sigma$ maintains the latest valuation of each variable.


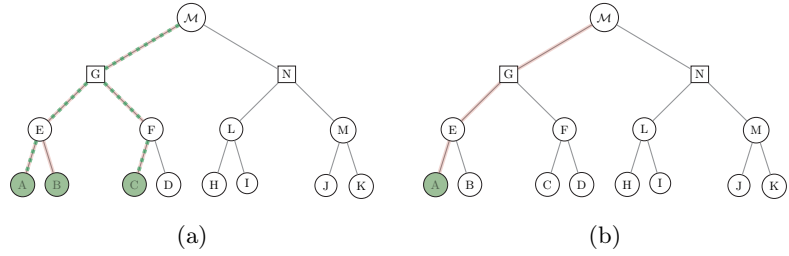
\begin{figure}
\begin{center}
\begin{tabular}{c@{}c}
\begin{minipage}{0.45\textwidth}
\centering
\resizebox{0.9\textwidth}{!}{
\begin{tikzpicture}
\node[circle, draw=Black](mo){$\mathcal{M}$};
\node[rectangle, draw=Black, below left=of mo, xshift=-1cm](g){G};
\node[rectangle, draw=Black, below right=of mo, xshift=1cm](n){N};
\node[circle, draw=Black, below left=of g](e){E};
\node[circle, draw=Black, below right=of g](f){F};
\node[circle, draw=Black, below left=of n](l){L};
\node[circle, draw=Black, below right=of n](m){M};
\node[circle, draw=Black, fill=OliveGreen, fill opacity=0.5, below left=of e, xshift=1cm](a){A};
\node[circle, draw=Black, fill=OliveGreen, fill opacity=0.5, below right=of e, xshift=-1cm](b){B};
\node[circle, draw=Black, fill=OliveGreen, fill opacity=0.5, below left=of f, xshift=1cm](c){C};
\node[circle, draw=Black,  below right=of f, xshift=-1cm](d){D};
\node[circle, draw=Black,  below left=of l, xshift=1cm](h){H};
\node[circle, draw=Black, below right=of l, xshift=-1cm](i){I};
\node[circle, draw=Black,  below left=of m, xshift=1cm](j){J};
\node[circle, draw=Black, below right=of m, xshift=-1cm](k){K};

\draw[-, Gray, thick](mo) -- (g);
\draw[-, Gray, thick](mo) -- (n);
\draw[-, Gray, thick](g) -- (e);
\draw[-, Gray, thick](g) -- (f);
\draw[-, Gray, thick](n) -- (l);
\draw[-, Gray, thick](n) -- (m);
\draw[-, Gray, thick](e) -- (a);
\draw[-, Gray, thick](e) -- (b);
\draw[-, Gray, thick](f) -- (c);
\draw[-, Gray, thick](f) -- (d);
\draw[-, Gray, thick](l) -- (h);
\draw[-, Gray, thick](l) -- (i);
\draw[-, Gray, thick](m) -- (j);
\draw[-, Gray, thick](m) -- (k);

\draw[-, BrickRed, line width=1mm, opacity=0.2](mo) -- (g);
\draw[-, BrickRed, line width=1mm, opacity=0.2](g) -- (e);
\draw[-, BrickRed, line width=1mm, opacity=0.2](g) -- (f);
\draw[-, BrickRed, line width=1mm, opacity=0.2](e) -- (a);
\draw[-, BrickRed, line width=1mm, opacity=0.2](e) -- (b);
\draw[-, BrickRed, line width=1mm, opacity=0.2](f) -- (c);

\draw[loosely dotted, Green, line width=1mm, opacity=0.5](mo) -- (g);
\draw[loosely dotted, Green, line width=1mm, opacity=0.5](g) -- (e);
\draw[loosely dotted, Green, line width=1mm, opacity=0.5](g) -- (f);
\draw[loosely dotted, Green, line width=1mm, opacity=0.5](e) -- (a);
\draw[loosely dotted, Green, line width=1mm, opacity=0.5](f) -- (c);

\end{tikzpicture}
}

\vspace{0.2cm}
(a)
\end{minipage}
&
\begin{minipage}{0.45\textwidth}
\centering
\resizebox{0.9\textwidth}{!}{
\begin{tikzpicture}
\node[circle, draw=Black](mo){$\mathcal{M}$};
\node[rectangle, draw=Black, below left=of mo, xshift=-1cm](g){G};
\node[rectangle, draw=Black, below right=of mo, xshift=1cm](n){N};
\node[circle, draw=Black, below left=of g](e){E};
\node[circle, draw=Black, below right=of g](f){F};
\node[circle, draw=Black, below left=of n](l){L};
\node[circle, draw=Black, below right=of n](m){M};
\node[circle, draw=Black, fill=OliveGreen, fill opacity=0.5, below left=of e, xshift=1cm](a){A};
\node[circle, draw=Black, below right=of e, xshift=-1cm](b){B};
\node[circle, draw=Black, below left=of f, xshift=1cm](c){C};
\node[circle, draw=Black, below right=of f, xshift=-1cm](d){D};
\node[circle, draw=Black, below left=of l, xshift=1cm](h){H};
\node[circle, draw=Black, below right=of l, xshift=-1cm](i){I};
\node[circle, draw=Black, below left=of m, xshift=1cm](j){J};
\node[circle, draw=Black, below right=of m, xshift=-1cm](k){K};

\draw[-, Gray, thick](mo) -- (g);
\draw[-, Gray, thick](mo) -- (n);
\draw[-, Gray, thick](g) -- (e);
\draw[-, Gray, thick](g) -- (f);
\draw[-, Gray, thick](n) -- (l);
\draw[-, Gray, thick](n) -- (m);
\draw[-, Gray, thick](e) -- (a);
\draw[-, Gray, thick](e) -- (b);
\draw[-, Gray, thick](f) -- (c);
\draw[-, Gray, thick](f) -- (d);
\draw[-, Gray, thick](l) -- (h);
\draw[-, Gray, thick](l) -- (i);
\draw[-, Gray, thick](m) -- (j);
\draw[-, Gray, thick](m) -- (k);

\draw[-, BrickRed, line width=1mm, opacity=0.2](mo) -- (g);
\draw[-, BrickRed, line width=1mm, opacity=0.2](g) -- (e);
\draw[-, BrickRed, line width=1mm, opacity=0.2](e) -- (a);
\end{tikzpicture}
}

\vspace{0.2cm}
(b)
\end{minipage}
\end{tabular}
\end{center}
\caption{Configuration and configuration state tree of statechart shown in Fig. ~\ref{f:constabl-example}: (a) A valid configuration state tree for \{A,C\} is marked by dotted green edges; An invalid configuration (composite state E has both its substates in the state tree) marked by red edges; (b) Another invalid configuration (shell state G does not have both its regions E and F in the state tree).}
\label{f:conf-code-example}
\end{figure}

\subsubsection{Configuration} (denoted by $\mathbb{C}$) is a set of \textit{atomic} states that are active at a given point. In 
a statechart model with only hierarchical composition, $\mathbb{C}$ is a singleton set. However, in concurrent models, $\mathbb{C}$ may have multiple atomic states. A configuration can also be represented by its configuration state tree(CST). $CST(\mathbb{C})$ is a state tree $st$ of statechart states such that each state in $\mathbb{C}$ is a leaf in $st$. All ancestors of all atomic states in $\mathbb{C}$ are contained in $st$. No other state is included as a node in $st$. Not all sets of atomic states are \emph{\textbf{valid configurations}}. A set $\mathbb{C}$ of atomic states is a valid configuration iff:
\begin{enumerate}
\item For each composite state $s \in st$, $s$ must have only one child in $st$ corresponding to one of its substates.
\item For each shell state $s \in st$, $s$ has a child node in $st$ corresponding to each of its regions.
\end{enumerate}

There are three major steps that happen as a part of a simulation step. In the first stages, the code that has to be executed during the simulation step is identified. In the second stage, the identified code is executed. In the third step, the configuration is changed from the source configuration to the destination configuration.
 
\subsection{Computing Transition Code}
We now present the details of the first stage, i.e. identification of the code to be execution or \emph{computation of transition code}.

\subsubsection{Enabled Transition.} In a given configuration $\mathbb{C}$, when an event $e$ has arrived, a transition $t$ is enabled when all three of the following conditions hold: 
\begin{enumerate}
\item $t.s \in CST(\mathbb{C})$, (i.e. $t$'s source state is one of the vertices of the configuration state tree of $\mathbb{C}$). 
\item $t.e = e$, (i.e. $t$'s trigger event is the same as the one that has arrived). 
\item $\sigma \vdash t.g \Downarrow true$, (i.e. in the given value environment $\sigma$, $t$'s guard evaluates to $true$). 
\end{enumerate}
It is possible that, for a given configuration $\mathbb{C}$ and event $e$, the number of enabled transitions is zero, one, or more than one. Hence, to perform the next step, we compute the set of enabled transitions $\mathcal{T}$. For example, in the example model in Fig. ~\ref{f:conf-code-example}, if $\mathbb{C}=\{A, C\}$ and $e=e_1$, then $\mathcal{T}=\{t_{AB}, t_{CD}\}$.

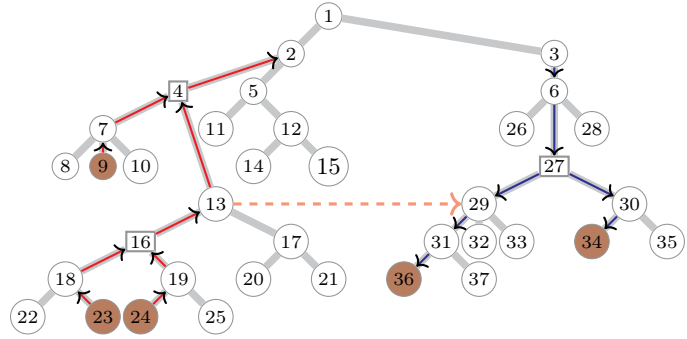
\begin{figure}
\begin{center}

\begin{tikzpicture}
\node[circle, draw=Gray, inner sep=0.05cm](1){\scriptsize 1};
\node[circle, draw=Gray, inner sep=0.05cm](2) at (-0.5, -0.5){\scriptsize 2};
\node[circle, draw=Gray, inner sep=0.05cm](3) at (3, -0.5){\scriptsize 3};
\node[rectangle, draw=Gray, thick, inner sep=0.05cm](4) at (-2, -1){\scriptsize 4};
\node[circle, draw=Gray, inner sep=0.05cm](5) at (-1, -1){\scriptsize 5};
\node[circle, draw=Gray, inner sep=0.05cm](6) at (3, -1){\scriptsize 6};
\node[circle, draw=Gray, inner sep=0.05cm](7) at (-3, -1.5){\scriptsize 7};
\node[circle, draw=Gray, inner sep=0.05cm](8) at (-3.5, -2){\scriptsize 8};
\node[circle, draw=Gray, fill=Brown!50, inner sep=0.05cm](9) at (-3, -2){\scriptsize 9};
\node[circle, draw=Gray, inner sep=0.05cm](10) at (-2.5, -2){\scriptsize 10};
\node[circle, draw=Gray, inner sep=0.05cm](11) at (-1.5, -1.5){\scriptsize 11};
\node[circle, draw=Gray, inner sep=0.05cm](12) at (-0.5, -1.5){\scriptsize 12};
\node[circle, draw=Gray, inner sep=0.05cm](13) at ([shift=({0.5cm,-1.5cm})]4){\scriptsize 13};
\node[circle, draw=Gray, inner sep=0.05cm](14) at ([shift=({-0.5cm,-0.5cm})]12){\scriptsize 14};
\node[circle, draw=Gray, inner sep=0.05cm](15) at ([shift=({0.5cm,-0.5cm})]12){15};
\node[rectangle, thick, draw=Gray, inner sep=0.05cm](16) at ([shift=({-1cm,-0.5cm})]13){\scriptsize 16};
\node[circle, draw=Gray, inner sep=0.05cm](17) at ([shift=({1cm,-0.5cm})]13){\scriptsize 17};
\node[circle, draw=Gray, inner sep=0.05cm](18) at ([shift=({-1cm,-0.5cm})]16){\scriptsize 18};
\node[circle, draw=Gray, inner sep=0.05cm](19) at ([shift=({0.5cm,-0.5cm})]16){\scriptsize 19};
\node[circle, draw=Gray, inner sep=0.05cm](20) at ([shift=({-0.5cm,-0.5cm})]17){\scriptsize 20};
\node[circle, draw=Gray, inner sep=0.05cm](21) at ([shift=({0.5cm,-0.5cm})]17){\scriptsize 21};
\node[circle, draw=Gray, inner sep=0.05cm](22) at ([shift=({-0.5cm,-0.5cm})]18){\scriptsize 22};
\node[circle, draw=Gray, fill=Brown!50, inner sep=0.05cm](23) at ([shift=({0.5cm,-0.5cm})]18){\scriptsize 23};
\node[circle, draw=Gray, fill=Brown!50, inner sep=0.05cm](24) at ([shift=({-0.5cm,-0.5cm})]19){\scriptsize 24};
\node[circle, draw=Gray, inner sep=0.05cm](25) at ([shift=({0.5cm,-0.5cm})]19){\scriptsize 25};

\node[circle, draw=Gray, inner sep=0.05cm](26) at ([shift=({-0.5cm,-0.5cm})]6){\scriptsize 26};
\node[rectangle, thick, draw=Gray, inner sep=0.05cm](27) at ([shift=({0cm,-1cm})]6){\scriptsize 27};
\node[circle, draw=Gray, inner sep=0.05cm](28) at ([shift=({0.5cm,-0.5cm})]6){\scriptsize 28};
\node[circle, draw=Gray, inner sep=0.05cm](29) at ([shift=({-1cm,-0.5cm})]27){\scriptsize 29};
\node[circle, draw=Gray, inner sep=0.05cm](30) at ([shift=({1cm,-0.5cm})]27){\scriptsize 30};
\node[circle, draw=Gray, inner sep=0.05cm](31) at ([shift=({-0.5cm,-0.5cm})]29){\scriptsize 31};
\node[circle, draw=Gray, inner sep=0.05cm](32) at ([shift=({0cm,-0.5cm})]29){\scriptsize 32};
\node[circle, draw=Gray, inner sep=0.05cm](33) at ([shift=({0.5cm,-0.5cm})]29){\scriptsize 33};
\node[circle, draw=Gray, fill=Brown!50, inner sep=0.05cm](34) at ([shift=({-0.5cm,-0.5cm})]30){\scriptsize 34};
\node[circle, draw=Gray, inner sep=0.05cm](35) at ([shift=({0.5cm,-0.5cm})]30){\scriptsize 35};
\node[circle, draw=Gray, fill=Brown!50, inner sep=0.05cm](36) at ([shift=({-0.5cm,-0.5cm})]31){\scriptsize 36};
\node[circle, draw=Gray, inner sep=0.05cm](37) at ([shift=({0.5cm,-0.5cm})]31){\scriptsize 37};

\draw[-, Gray!50, line width=1mm](1) -- (2);
\draw[-, Gray!50, line width=1mm](1) -- (3);
\draw[-, Gray!50, line width=1mm](2) -- (4);
\draw[-, Gray!50, line width=1mm](2) -- (5);
\draw[-, Gray!50, line width=1mm](4) -- (7);
\draw[-, Gray!50, line width=1mm](4) -- (13);
\draw[-, Gray!50, line width=1mm](5) -- (11);
\draw[-, Gray!50, line width=1mm](5) -- (12);
\draw[-, Gray!50, line width=1mm](7) -- (8);
\draw[-, Gray!50, line width=1mm](7) -- (9);
\draw[-, Gray!50, line width=1mm](7) -- (10);
\draw[-, Gray!50, line width=1mm](12) -- (14);
\draw[-, Gray!50, line width=1mm](12) -- (15);
\draw[-, Gray!50, line width=1mm](13) -- (16);
\draw[-, Gray!50, line width=1mm](13) -- (17);
\draw[-, Gray!50, line width=1mm](16) -- (18);
\draw[-, Gray!50, line width=1mm](16) -- (19);
\draw[-, Gray!50, line width=1mm](17) -- (20);
\draw[-, Gray!50, line width=1mm](17) -- (21);
\draw[-, Gray!50, line width=1mm](18) -- (22);
\draw[-, Gray!50, line width=1mm](18) -- (23);
\draw[-, Gray!50, line width=1mm](19) -- (24);
\draw[-, Gray!50, line width=1mm](19) -- (25);

\draw[-, Gray!50, line width=1mm](3) -- (6);
\draw[-, Gray!50, line width=1mm](6) -- (26);
\draw[-, Gray!50, line width=1mm](6) -- (27);
\draw[-, Gray!50, line width=1mm](6) -- (28);
\draw[-, Gray!50, line width=1mm](27) -- (29);
\draw[-, Gray!50, line width=1mm](27) -- (30);
\draw[-, Gray!50, line width=1mm](29) -- (31);
\draw[-, Gray!50, line width=1mm](29) -- (32);
\draw[-, Gray!50, line width=1mm](29) -- (33);
\draw[-, Gray!50, line width=1mm](30) -- (34);
\draw[-, Gray!50, line width=1mm](30) -- (35);
\draw[-, Gray!50, line width=1mm](31) -- (36);
\draw[-, Gray!50, line width=1mm](31) -- (37);

\draw[->, draw=Red!50, dashed, very thick](13) -- (29);

\draw[->, draw=Red, thick](4) -- (2);
\draw[->, draw=Red, thick](7) -- (4);
\draw[->, draw=Red, thick](13) -- (4);
\draw[->, draw=Red, thick](9) -- (7);
\draw[->, draw=Red, thick](16) -- (13);
\draw[->, draw=Red, thick](18) -- (16);
\draw[->, draw=Red, thick](19) -- (16);
\draw[->, draw=Red, thick](23) -- (18);
\draw[->, draw=Red, thick](24) -- (19);

\draw[->, draw=Blue, thick](3) -- (6);
\draw[->, draw=Blue, thick](6) -- (27);
\draw[->, draw=Blue, thick](27) -- (29);
\draw[->, draw=Blue, thick](27) -- (30);
\draw[->, draw=Blue, thick](29) -- (31);
\draw[->, draw=Blue, thick](31) -- (36);
\draw[->, draw=Blue, thick](30) -- (34);
\end{tikzpicture}
\end{center}
\caption{Transition state trees}
\label{f:state-tree}
\end{figure}

We define a constructor $\textsc{tree}: node \times tree set \rightarrow tree$ as function such that $\textsc{tree}(n, S)$ creates a tree rooted at node $n$ with all trees is set $S$ as subtrees of $n$.

\textbf{Subtree}. The subtree of a node $n$, given by $\mathbb{T}(n)$, is given by 

\begin{align*}
\mathbb{T}(n) = \textsc{tree}(n, \{\mathbb{T}(c) \forall c \in childnodes(n)\})
\end{align*}

\subsubsection{Sliced subtree ($\widehat{\mathbb{T}}$).} For any node $n$, the sliced subtree $\widehat{tr} = \widehat{\mathbb{T}}(n, C)$ w.r.t. to node set $C$ is a subtree of $n$ such that each path in $\widehat{tr}$ from its root goes to a member of $C$. Here, $C$ is a set of nodes that contains some of the leaves of $\mathbb{T}(n)$ possibly along with other nodes not in $\mathbb{T}(n)$.

\subsubsection{Initial subtree ($ST_i$).} For any node $n$, the initial subtree $ST_i(n)$ is given by the following:

\begin{equation*}
\begin{split}
ST_i(n) = 
& \textsc{tree}(n, \{ST_i(n.I)\}) \textbf{ if } n.\tau \in \{Composite, Statechart\} \\
& \textsc{tree}(n, \{\}) \textbf{ if } n.\tau = Atomic \\
& \textsc{tree}(n, \{ST_i(c),\ \forall c | c \in childnodes(n)\}) \textbf{ if } n.\tau = Shell
\end{split}
\end{equation*}

\subsubsection{Transition state tree.} When an enabled transition is fired, a collection of code blocks get executed. They are called \emph{source side code} which corresponds to the exit blocks of the states in source configuration state tree, the action code block of the transition and \emph{destination side code} which are the entry code blocks of the states in destination configuration state tree, in that order. In a non-concurrent case, both the source side code and the destination side code are a sequence of code blocks. However, in a concurrent case, the code blocks are non-sequential, and in general, can be viewed as two trees corresponding to the containment hierarchy of the states that they are a part of. These two trees are in turn called the \emph{source side state tree} ($ST_s(t, \mathbb{C})$) and the \emph{destination side state tree} ($ST_d(t)$) of a transition being fired in a particular configuration. The direction of the arrows are along the general flow of control of the code execution during the simulation step. $ST_s$ and $ST_d$ are defined as follows:

\begin{align*}
ST_s(t, \mathbb{C}) &= \\
& \textbf{let}\ l = t.src \sqcup t.dest \textbf{ in} \\
& \textbf{let}\ s =\\
&\hspace{1cm}\textbf{if } t.src \prec_1 l \textbf{ then } t.src \\
&\hspace{1cm}\textbf{else } \textbf{such that } s \prec_1 l \land t.src \prec_* s \textbf{ in}\\
& \widehat{\mathbb{T}}(s,  \mathbb{C})
\end{align*}

\begin{align*}
ST_d(t) &= \\
& \textbf{let}\ s =\\
&\hspace{1cm}\textbf{if } t.dest \prec_1 l \textbf{ then } t.dest \\
&\hspace{1cm}\textbf{else } \textbf{such that } s \prec_1 l \land t.dest \prec_* s \textbf{ in}\\
&ST_i(s)
\end{align*}

\begin{example}
Figure~\ref{f:state-tree} shows a state containment hierarchy. We have used rectangular boxes to indicate shell states, and circular boxes to show other types of states. The red dashed line from state 13 to state 29 shown an enabled transition $t$ which is fired. The current configuration is $\mathbb{C}=\{9, 23, 24, 34, 36\}$ and the corresponding states are shown filled with brown colour. State $1 = 13 \sqcup 29$. Hence, state 2 is the last state to be exited on the source side and state 3 is the first state to enterred on the destination side.
The source side state tree $ST_s(t, \mathbb{C})$ and the destination side state tree $ST_d(t)$ are highlighted in red and blue edges respectively. \qed
\end{example}

\subsubsection{Control Flow Graph Tree.} \label{s:cfgtree}

As mentioned above, the source side code tree for an enabled transition $t$ in a configuration $\mathbb{C}$ is the tree of exit code blocks of the corresponding source side state tree $ST_s(t,\mathbb{C})$ of $t$. Similarly, the destination side code tree for an enabled transition $t$ in a configuration $\mathbb{C}$ is the tree of entry code blocks of the corresponding destination side state tree $ST_s(T)$ of $t$. Mathematically:

\begin{align*} \label{e:conflict}
CFGTree_s(t, \mathbb{C}) = treemap(s \rightarrow \textsc{cfg}(s.\mathcal{X}), ST_s(t, \mathbb{C})) \\ 
CFGTree_d(t) = treemap(s \rightarrow \textsc{cfg}([init(s.V),s.\mathcal{N}]), ST_d(t)) 
\end{align*}

Here, the constructor \textsc{cfg} takes a code block (in the form of its abstract syntax tree) and gives its control flow graph (CFG). Control flow graphs are presented in detail in Section~\ref{s:cfg} where it will be needed to discuss code execution. Here, it suffices to proceed with an informal understanding of control flow graphs. $treemap: (\alpha \rightarrow \beta) \times \alpha\ tree \rightarrow \beta\ tree$ is a function such that $treemap(f, t)$ gives a tree $t'$ that is identical in shape as the $t$ except that the value on each node of $t'$ is $v' = f(v)$ where $v$ is the value on the corresponding node of $t$. While computing the source side CFG tree $CFGTree_s(t, \mathbb{C})$, we include the CFGs of the exit code of each state in the source side state tree $ST_s(t, \mathbb{C})$. While computing the destination side CFG tree $CFGTree_d(t)$, we include the CFGs of the concatenation of the variable initialisation code and entry code of each state in the destination side state tree $ST_d(t)$.

\subsubsection{Code.} The code that gets executed during a simulation step is represented by the following data-type $Code$ as follows:

\begin{align*}
Code &= Seq([c_1, c_2, ..., c_n]) \text{ where }c_1, c_2, ..., c_n : Code \\
            &= Conc(\{c_1, c_2, ..., c_n\}) \text{ where }c_1, c_2, ..., c_n : Code \\
            &= CFGCode(cfg) \text{ where $cfg = \textsc{cfg}(b)$ and $b$ a code block} \\
\end{align*}

Henceforth, we abbreviate $Seq([c_1, c_2, ..., c_n])$ as $\langle c_1, c_2, ... c_n \rangle$ and $Conc(\{c_1, c_2, ..., c_n\})$ as $[c_1 | c_2 | ...| c_n]$.

\begin{example}
In Figure~\ref{f:conf-code-example}, when transition $t_{GN}$ is fired in configuration $\{A, C\}$, the code that gets executed is: 
\begin{align*}
\mathcal{C}(t_{GN}) &= Seq([Conc(\{Seq([A.\mathcal{X}, E.\mathcal{X}]), Seq([C.\mathcal{X}, F.\mathcal{X}])\}), G.\mathcal{X}, \\
            & t_{GN}.a, \\
            &  N.\mathcal{N}, Conc(\{Seq([N.\mathcal{N}, H.\mathcal{N}]), Seq([M.\mathcal{N}, J.\mathcal{N}])\})]) \\
\end{align*}

In abbreviated form, the above is written as: 
\begin{align*}
\mathcal{C}(t_{GN}) &= \langle [\langle A.\mathcal{X}, E.\mathcal{X}\rangle | \langle C.\mathcal{X}, F.\mathcal{X} \rangle ], G.\mathcal{X}, t_{GN}.a, N.\mathcal{N}, [\langle L.\mathcal{N}, H.\mathcal{N} \rangle | \langle M.\mathcal{N}, J.\mathcal{N}\rangle] \rangle
\end{align*}
\qed
\end{example}

The function $\mathcal{c}: StateTree \rightarrow Code$ takes a State tree $st$, and gives its code as follows: if $st$ is source tree then $s.cfg \rightarrow s.\mathcal{X}$ and if $st$ is destination tree then $s.cfg \rightarrow [init(s.\mathcal{V}_l), s.\mathcal{N}]$.

\begin{align*}
\mathcal{c}(st) &= CFGCode(s.cfg) \text{ \textbf{if} } st=\textsc{tree}(s,\{\}) \\
& \hspace{0.5cm}Seq([CFGCode(s.cfg),\ \mathcal{C}(c)]) \text{ \textbf{if} } st=\textsc{tree}(s,\{c\}) \\
& \hspace{0.5cm}Seq([CFGCode(s.cfg), \ Conc(\{\mathcal{C}(st_1), \mathcal{C}(st_2), ..., \mathcal{C}(st_n)\})]) \\
& \hspace{1cm} \text{ \textbf{if} } st=\textsc{tree}(s,\{c_1, c_2, ..., c_n\}) 
\end{align*}


\subsubsection{Transition Code}
The code to be executed with a transition $t$ gets fired in a configuration $\mathbb{C}$ is given by:
\begin{align*}
Code(t, \mathbb{C}) &= Seq([\mathcal{C}_s(t, \mathbb{C})\ ,\textsc{cfg}(t.a)\ ,\mathcal{C}_d(t, \mathbb{C})])\\ 
&\hspace{0.5cm}\textbf{ where, } \mathcal{C}_s(t, \mathbb{C}) = \mathcal{C}(CFGTree_s(t,\mathbb{C}))\textbf{ and }\\
&\hspace{0.5cm} \mathcal{C}_d(t, \mathbb{C}) = rev(CFGTree_d(t))\textbf{ and }\\
&\hspace{0.5cm}CFGTree_s(t, \mathbb{C}) = treemap(s \rightarrow \textsc{cfg}(s.\mathcal{X}), ST_s(t, \mathbb{C})) \textbf{ and }\\ 
&\hspace{0.5cm}CFGTree_d(t) = treemap(s \rightarrow \textsc{cfg}(Seq([init(s.\mathcal{V}_l),s.\mathcal{N}]), ST_d(t))\\
\end{align*}

Note the code reversal done to compute $\mathcal{C}_d(t, \mathbb{C})$. Code reversal function done to a code $c : Code$ is defined as follows:
\begin{align*}
rev(c) &= \\
& c \text{ \textbf{if} } c \text{ \textbf{is} } CFGCode \\
& \langle rev(c_n), rev(c_{n - 1}) ..., rev(c_2), rev(c_1) \rangle \text{ \textbf{if} } c \text{ \textbf{is} } \langle c_1, c_2, ..., c_n \rangle \\
& [rev(c_1) | rev(c_2) | ... | rev(c_{n-1}) | rev(c_n)] \text{ \textbf{if} } c \text{ \textbf{is} } [c_1 | c_2 | ... | c_n] \\
\end{align*}

Finally, the entire code to be executed during a simulation step is the concurrent composition of the codes of the enabled transitions. For the set of enabled transitions $\mathcal{T} = \{t_1, t_2, ..., t_n\}$, code is given by:

\begin{align*}
\mathcal{C}(\mathcal{T}) =  [\mathcal{C}(t_1, \mathbb{C}) | \mathcal{C}(t_2, \mathbb{C}), ... | \mathcal{C}(t_n, \mathbb{C})]
\end{align*}

\begin{figure}
\begin{center}
\begin{tabular}{c @{} c}
\begin{minipage}{0.45\textwidth}
\begin{center}
\begin{tikzpicture}
\node[rectangle, draw=Black, minimum size=1cm, rounded corners](a){};
\node[below right=0.1cm of a.north west](la){A};
\node[rectangle, draw=Black, rounded corners, inner sep=1cm, fit=(a)](b){};
\node[below right=0.1cm of b.north west](lb){B};

\draw[->, Red, thick](b.30) --node[above]{{\color{Black}$t_1: e[g]$}}+(2cm,0);
\draw[->, Red, thick](a.330) --node[above, near end]{{\color{Black}$t_2: e[g]$}}+(3cm,0);
\end{tikzpicture}

\vspace{0.5cm}
(a)
\end{center}
\end{minipage}
&
\begin{minipage}{0.45\textwidth}
\begin{center}
\begin{tikzpicture}
\node[rectangle, draw=Black, minimum size=1cm, rounded corners](a){};
\node[below right=0.1cm of a.north west](la){A};
\node[rectangle, draw=Black, minimum size=1cm, rounded corners, below=of a](b){};
\node[below right=0.1cm of b.north west](lb){B};
\node[rectangle, draw=Black, rounded corners, inner sep=0.5cm, fit=(a)(b)](c){};
\node[below right=0.1cm of c.north west](lc){C};
\draw[-, dashed](c.west) -- (c.east);

\draw[->, Red, thick](a) --node[above, near end]{{\color{Black}$t_2: e[g]$}}+(2.5cm,0);
\draw[->, Red, thick](b) --node[above, near end]{{\color{Black}$t_1: e[g]$}}+(2.5cm,0);
\end{tikzpicture}

\vspace{0.5cm}
(b)
\end{center}
\end{minipage}
\end{tabular}
\end{center}
\caption{Conflicting Transitions: (a) Both $t_1$ and $t_2$ would cause A.$\mathcal{X}$ and B.$\mathcal{X}$ to execute; (b) Both $t_1$ and $t_2$ would cause C.$\mathcal{X}$ to execute.}
\label{f:conflict}
\end{figure}
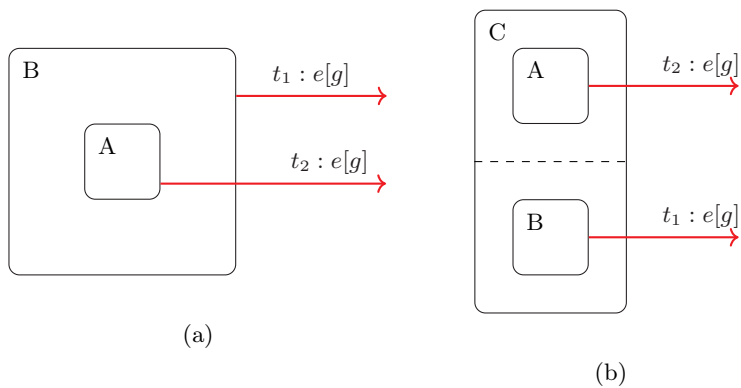

\subsubsection{Conflicting transitions.} For a given configuration $\mathbb{C}$ and event $e$, two enabled transitions are said to \emph{conflict} if, fired concurrently, they would lead to repeated execution of some code block. 

\begin{align*}
\sigma, \mathbb{C} & \vdash \forall t_1, t_2 \in \mathcal{T}, \\
& \hspace{1cm}conflict(t_1, t_2) = true \text{ if } \mathcal{C}(t_1, \mathbb{C}) \cap \mathcal{C}(t_2, \mathbb{C}) \neq \phi 
\end{align*}

In the above, we use the set intersection ($\cap$) to indicate the intersection between the code blocks in $\mathcal{C}(t_1, \mathbb{C})$ and $\mathcal{C}(t_2, \mathbb{C})$.

\begin{example}[Conflicting transitions] \label{ex:conflict}
Figure~\ref{f:conflict} shows two scenarios in which transitions may conflict. Figure~\ref{f:conflict}(a) shows a case in a sequential (non-concurrent) scenario where two enabled transitions $t_1$ and $t_2$ emerge from two states which are in ancestor-descendent relationship. This case which we call \emph{non-determinism}, makes it impossible to determine which transition to take, unless we use an external conflict resolution rule (e.g. clockwise arrangement \cite{stateflow}, outside-in \cite{710975} etc). We consider such conflict resolution undesirable as it may sneak in undesirable behaviour without the knowledge of the modelling engineer. Instead, such cases should be flagged out wherever possible, and the engineer should be given the choice to resolve in a way as desirable to him/her. Figure~\ref{f:conflict}(b) shows a case of a concurrent model. Two enabled transitions emerge out of the two regions A and B of a shell state C. Both transitions would lead to the execution of C.$\mathcal{X}$, which would be invalid.
\end{example}

\subsubsection{Valid simulation step.}

In any simulation step, any code block must execute at most once. Therefore, conflicting transitions are not allowed in a valid simulation.

\subsection{Executing Transition Code}
After the transition code $\mathcal{C}(t, \mathbb{C})$ is computed, the simulator will proceed to execute it. In this subsection, we present the details of this process.

Each step in code execution involves interpretation of single statements in the code. The code can be concurrent. Unlike many existing implementations of Statecharts \cite{stateflow}\cite{harel_statemate_1996}\cite{vondeerbeek}\cite{chsm}\cite{yakindu}\cite{Decan2020}, we allow arbitrary interleaving between concurrently executing code. We assume individual instructions to execute atomically. Though finer grained interleaving is possible in principle, we consider that not to be necessary feature at the statechart modelling level.

\begin{example}[Transition Code] \label{ex:tc}
Consider the model shown in Figure~\ref{f:conf-code-example}. Source configuration $\mathbb{C} = \{A, C\}$. For this example, assume that the code blocks in the model are as follows:
\begin{itemize}
\item \textbf{Entry codes:} $A.\mathcal{N} = \{i_{A\mathcal{N}_1};i_{A\mathcal{N}_2};\}$, $B.\mathcal{N} = \{i_{B\mathcal{N}_1};i_{B\mathcal{N}_2};\}$ etc.
\item \textbf{Exit codes:} $A.\mathcal{X} = \{i_{A\mathcal{X}_1};i_{A\mathcal{X}_2};\}$, $B.\mathcal{X} = \{i_{B\mathcal{X}_1};i_{B\mathcal{X}_2};\}$ etc.
\item \textbf{Transition action codes:} $t_{AB} = \{i_{AB_1};i_{AB_2};\}$, $t_{CD} = \{i_{CD_1};i_{CD_2};\}$ etc.
\end{itemize} 

\begin{enumerate}
\item \textbf{Case 1 (Concurrent and disjoint).} Suppose, the input event is $e_1$. The code to be executed in this case is $\{\mathcal{C}(t_{AB}, \mathbb{C}), \mathcal{C}(t_{CD}, \mathbb{C})\}$ = $\{A.\mathcal{X}\ t_{AB}.a\ B.\mathcal{N}, C.\mathcal{X}\ t_{CD}.a\ D.\mathcal{N}\}$. A possible trace generated when the above code executes: $i_{A\mathcal{X}_1}$, $i_{C\mathcal{X}_1}$, $i_{A\mathcal{X}_2}$, $i_{C\mathcal{X}_2}$, $i_{AB_1}$, $i_{CD_1}$, $i_{AB_2}$, $i_{CD_2}$, $i_{B\mathcal{N}_1}$, $i_{D\mathcal{N}_1}$, $i_{B\mathcal{N}_2}$, $i_{D\mathcal{N}_2}$.
\item \textbf{Case 2 (Concurrent, joining and forking).} Suppose, the input event is $e$. The code to be executed is

$\mathcal{C}(t_{GN}, \mathbb{C})$ = $\langle [\langle A.\mathcal{X}, E.\mathcal{X}\rangle | \langle C.\mathcal{X}, F.\mathcal{X} \rangle ], G.\mathcal{X}, t_{GN}.a, N.\mathcal{N},$ \\ $ [\langle L.\mathcal{N}, H.\mathcal{N} \rangle | \langle M.\mathcal{N}, J.\mathcal{N}\rangle] \rangle$. A possible trace generated when the \\ above code executes: $i_{A\mathcal{X}_1}$, $i_{C\mathcal{X}_1}$, $i_{A\mathcal{X}_2}$, $i_{C\mathcal{X}_2}$, $i_{E\mathcal{X}_1}$, $i_{F\mathcal{X}_1}$, $i_{E\mathcal{X}_2}$, $i_{F\mathcal{X}_2}$, $i_{G\mathcal{X}_1}$, $i_{G\mathcal{X}_2}$, $i_{GN_1}$, $i_{GN_2}$, $i_{N\mathcal{N}_1}$, $i_{N\mathcal{N}_2}$, $i_{L\mathcal{N}_1}$, $i_{M\mathcal{N}_1}$,  $i_{L\mathcal{N}_2}$, $i_{M\mathcal{N}_2}$, $i_{H\mathcal{N}_1}$, $i_{J\mathcal{N}_1}$, $i_{H\mathcal{N}_2}$, $i_{J\mathcal{N}_2}$.
\end{enumerate}
\qed
\end{example}

\subsubsection{Code Containment Tree.} \label{s:cct}
As is clear, the \emph{code} type is recursive and can represent a hierarchical arrangement of code blocks. We call this hierarchy as code hierarchy and represent the same by a tree called \emph{code containment tree}. Sequence code and concurrent code form the internal nodes of this tree and CFG code form the leaf nodes of this tree. This tree is useful in navigating the code during code execution. Please note that this is different from the code tree mentioned in Section~\ref{s:cfgtree} which mimics the state hierarchy.

\begin{example}[Code containment tree] \label{ex:cct}

The code shown in Example~\ref{ex:tc} are pictorially illustrated in the form of code containment trees in Figure~\ref{f:cct}. We use rectangular nodes with pointed vertices to show a concurrent code, rectangular nodes with rounded corners to show a sequential code and an unbordered node to show CFG codes. 
\qed
\end{example}

\begin{figure}
\begin{tabular}{c @{} c}
\begin{minipage}{0.5\textwidth}
\begin{center}
\scalebox{0.75}{
\begin{tikzpicture}
\node[rectangle, draw, thick](Ce1C){$\mathcal{C}(e_1, \mathbb{C})$};

\node[rectangle, rounded corners, draw, below left=1cm of Ce1C](Cab){$\mathcal{C}(t_{AB}, e_1)$};
\node at ([shift=({-1cm,-1cm})]Cab) (Ax){$A.\mathcal{X}$};
\node at ([shift=({0cm,-1cm})]Cab) (taba){$t_{AB}.a$};
\node at ([shift=({1cm,-1cm})]Cab) (Bn){$B.\mathcal{N}$};

\node[rectangle, rounded corners, draw, below right=1cm of Ce1C](Ccd){$\mathcal{C}(t_{AB}, e_1)$};
\node at ([shift=({-1cm,-1cm})]Ccd) (Cx){$C.\mathcal{X}$};
\node at ([shift=({0cm,-1cm})]Ccd) (tcda){$t_{CD}.a$};
\node at ([shift=({1cm,-1cm})]Ccd) (Dn){$D.\mathcal{N}$};

\draw[-, thick](Ce1C) -- (Cab);
\draw[-, thick](Ce1C) -- (Ccd);
\draw[-, thick](Cab) -- (Ax);
\draw[-, thick](Cab) -- (taba);
\draw[-, thick](Cab) -- (Bn);
\draw[-, thick](Ccd) -- (Cx);
\draw[-, thick](Ccd) -- (tcda);
\draw[-, thick](Ccd) -- (Dn);

\end{tikzpicture}
}
\end{center}
\end{minipage}
&
\begin{minipage}{0.5\textwidth}
\begin{center}
\scalebox{0.75}{
\begin{tikzpicture}
\node[rectangle, rounded corners, draw, thick](CeC){$\mathcal{C}(e, \mathbb{C})$};
\node [rectangle, draw, thick] at ([shift=({-2cm,-1cm})]CeC) (c1){};
\node at ([shift=({-1cm,-1cm})]CeC) (Gx){$G.\mathcal{X}$};
\node at ([shift=({0cm,-1cm})]CeC) (tgna){$t_{GN}.a$};
\node at ([shift=({1cm,-1cm})]CeC) (Nn){$N.\mathcal{N}$};
\node [rectangle, draw, thick] at ([shift=({2cm,-1cm})]CeC) (c2){};

\node[rectangle, rounded corners, draw, thick](s1) at ([shift=({-1cm,-1cm})]c1){};
\node[rectangle, rounded corners, draw, thick](s2) at ([shift=({1cm,-1cm})]c1){};
\node[rectangle, rounded corners, draw, thick](s3) at ([shift=({-1cm,-1cm})]c2){};
\node[rectangle, rounded corners, draw, thick](s4) at ([shift=({1cm,-1cm})]c2){};

\node at ([shift=({-0.5cm,-1cm})]s1) (Ax){$A.\mathcal{X}$};
\node at ([shift=({0.5cm,-1cm})]s1) (Ex){$E.\mathcal{X}$};
\node at ([shift=({-0.5cm,-1cm})]s2) (Cx){$C.\mathcal{X}$};
\node at ([shift=({0.5cm,-1cm})]s2) (Fx){$F.\mathcal{X}$};
\node at ([shift=({-0.5cm,-1cm})]s3) (Ln){$L.\mathcal{N}$};
\node at ([shift=({0.5cm,-1cm})]s3) (Hn){$H.\mathcal{N}$};
\node at ([shift=({-0.5cm,-1cm})]s4) (Mn){$M.\mathcal{N}$};
\node at ([shift=({0.5cm,-1cm})]s4) (Jn){$J.\mathcal{N}$};

\draw[-, thick](CeC) -- (c1);
\draw[-, thick](CeC) -- (Gx);
\draw[-, thick](CeC) -- (tgna);
\draw[-, thick](CeC) -- (Nn);
\draw[-, thick](CeC) -- (c2);

\draw[-, thick](c1) -- (s1);
\draw[-, thick](c1) -- (s2);
\draw[-, thick](c2) -- (s3);
\draw[-, thick](c2) -- (s4);

\draw[-, thick](s1) -- (Ax);
\draw[-, thick](s1) -- (Ex);
\draw[-, thick](s2) -- (Cx);
\draw[-, thick](s2) -- (Fx);
\draw[-, thick](s3) -- (Ln);
\draw[-, thick](s3) -- (Hn);
\draw[-, thick](s4) -- (Mn);
\draw[-, thick](s4) -- (Jn);

\end{tikzpicture}
}
\vspace{0.5cm}
\end{center}
\end{minipage}
\\
(a) & (b)
\end{tabular}

\caption{Example: Code containment tree of the code in Example~\ref{ex:tc}.}
\label{f:cct}
\end{figure}
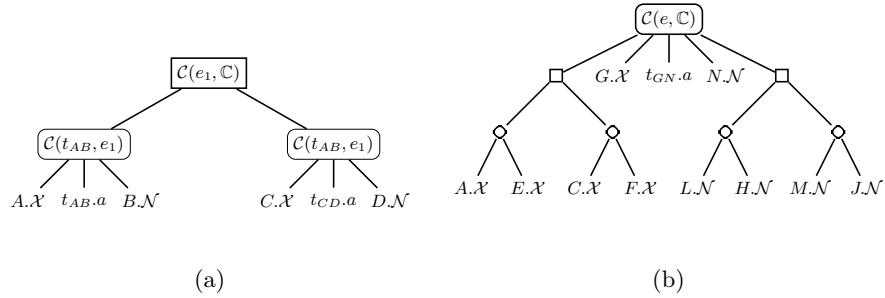

\begin{figure}
\begin{center}
\resizebox{\textwidth}{!}{
\begin{tikzpicture}
\node[rectangle, draw, fill](1){};
\node[above=0.1cm of 1](l1){\scriptsize 1};

\node[rectangle, draw, fill=Gray!30, minimum size=1cm, right=0.2cm of 1](Cx){$C.\mathcal{X}$};
\node[rectangle, draw, fill, right=0.2cm of Cx](3){};
\node[above=0.1cm of 3](l3){\scriptsize 3};

\node[rectangle, draw, fill=Gray!30, minimum size=1cm, right=0.2cm of 3](Fx){$F.\mathcal{X}$};

\node[rectangle, draw, fill, below= of 1](2){};
\node[above=0.1cm of 2](l2){\scriptsize 2};

\node[rectangle, draw, fill=Gray!30, minimum size=1cm, right=0.2cm of 2](Ax){$A.\mathcal{X}$};
\node[rectangle, draw, fill, right=0.2cm of Ax](4){};
\node[above=0.1cm of 4](l4){\scriptsize 4};

\node[rectangle, draw, fill=Gray!30, minimum size=1cm, right=0.2cm of 4](Ex){$E.\mathcal{X}$};

\node[rectangle, draw, fill, right=0.3cm of Fx, yshift=-0.6cm](5){};
\node[above=0.1cm of 5](l5){\scriptsize 5};

\node[rectangle, draw, fill=Gray!30, minimum size=1cm, right=0.2cm of 5](Gx){$G.\mathcal{X}$};
\node[rectangle, draw, fill, right=0.2cm of Gx](6){};
\node[above=0.1cm of 6](l6){\scriptsize 6};

\node[rectangle, draw, fill=Gray!30, minimum size=1cm, right=0.2cm of 6](ta){$t_{GN}.a$};
\node[rectangle, draw, fill, right=0.2cm of ta](7){};
\node[above=0.1cm of 7](l7){\scriptsize 7};

\node[rectangle, draw, fill=Gray!30, minimum size=1cm, right=0.2cm of 7](Nn){$N.\mathcal{N}$};
\node[rectangle, draw, fill, right=0.2cm of Nn](8){};
\node[above=0.1cm of 8](l8){\scriptsize 8};

\node[rectangle, draw, fill=Gray!30, minimum size=1cm, right=0.3cm of 8, yshift=0.6cm](Ln){$L.\mathcal{N}$};
\node[rectangle, draw, fill, right=0.2cm of Ln](9){};
\node[above=0.1cm of 9](l9){\scriptsize 9};

\node[rectangle, draw, fill=Gray!30, minimum size=1cm, right=0.2cm of 9](Hn){$H.\mathcal{N}$};

\node[rectangle, draw, fill=Gray!30, minimum size=1cm, below=0.2cm of Ln](Mn){$M.\mathcal{N}$};
\node[rectangle, draw, fill, right=0.2cm of Mn](10){};
\node[above=0.1cm of 10](l10){\scriptsize 10};

\node[rectangle, draw, fill=Gray!30, minimum size=1cm, right=0.2cm of 10](Jn){$J.\mathcal{N}$};

\node[rectangle, draw, dashed, inner sep=0.3cm, fit=(l1) (Jn)](border1){};

\draw[->, Red, thick](1) -- (Cx);
\draw[->, Red, thick](Cx) -- (3);
\draw[->, Red, thick](3) -- (Fx);
\draw[->, Red, thick](Fx) -- (5);

\draw[->, Red, thick](2) -- (Ax);
\draw[->, Red, thick](Ax) -- (4);
\draw[->, Red, thick](4) -- (Ex);
\draw[->, Red, thick](Ex) -- (5);
\draw[->, Red, thick](5) -- (Gx);
\draw[->, Red, thick](Gx) -- (6);
\draw[->, Red, thick](6) -- (ta);
\draw[->, Red, thick](ta) -- (7);
\draw[->, Red, thick](7) -- (Nn);
\draw[->, Red, thick](Nn) -- (8);
\draw[->, Red, thick](8) -- (Ln);
\draw[->, Red, thick](8) -- (Mn);
\draw[->, Red, thick](Ln) -- (9);
\draw[->, Red, thick](9) -- (Hn);
\draw[->, Red, thick](Mn) -- (10);
\draw[->, Red, thick](10) -- (Jn);
\end{tikzpicture}
}

\vspace{0.2cm}
(a)
\vspace{0.5cm}

\begin{tikzpicture}
\node[rectangle, draw, fill](1){};
\node[above=0.1cm of 1](l1){\scriptsize 1};
\node[rectangle, draw, fill=White, right=0.3cm of 1](icx1){$i_{C\mathcal{X}_1}$};
\node[rectangle, draw, fill, right=0.2cm of icx1](1_1){};
\node[above=0.1cm of 1_1](l1_1){\scriptsize 1.1};
\node[rectangle, draw, fill=White, right=0.2cm of 1_1](icx2){$i_{C\mathcal{X}_2}$};
\node[rectangle, draw, fill, right=0.3cm of icx2](3){};
\node[above=0.1cm of 3](l3){\scriptsize 3};

\begin{pgfonlayer}{background}
   \node [draw, fill=Gray!30,fit=(icx1) (icx2) (l1_1)] (Cx){};
\end{pgfonlayer}
\node[above=0.1cm of Cx](lCx){\scriptsize $C.\mathcal{X}$};

\node[rectangle, draw, fill=White, right=0.3cm of 3](ifx1){$i_{F\mathcal{X}_1}$};
\node[rectangle, draw, fill, right=0.2cm of ifx1](3_1){};
\node[above=0.1cm of 3_1](l3_1){\scriptsize 3.1};
\node[rectangle, draw, fill=White, right=0.2cm of 3_1](ifx2){$i_{F\mathcal{X}_2}$};

\begin{pgfonlayer}{background}
   \node [draw, fill=Gray!30,fit=(ifx1) (ifx2) (l3_1)] (Fx){};
\end{pgfonlayer}
\node[above=0.1cm of Fx](lFx){\scriptsize $F.\mathcal{X}$};

\node[rectangle, draw, fill, below=of 1](2){};
\node[above=0.1cm of 2](l2){\scriptsize 2};
\node[rectangle, draw, fill=White, right=0.3cm of 2](iax1){$i_{A\mathcal{X}_1}$};
\node[rectangle, draw, fill, right=0.2cm of iax1](2_1){};
\node[above=0.1cm of 2_1](l2_1){\scriptsize 2.1};
\node[rectangle, draw, fill=White, right=0.2cm of 2_1](iax2){$i_{A\mathcal{X}_2}$};
\node[rectangle, draw, fill, right=0.3cm of iax2](4){};
\node[above=0.1cm of 4](l4){\scriptsize 4};

\begin{pgfonlayer}{background}
   \node [draw, fill=Gray!30,fit=(iax1) (iax2) (l2_1)] (Ax){};
\end{pgfonlayer}
\node[below=0.1cm of Ax](lAx){\scriptsize $A.\mathcal{X}$};

\node[rectangle, draw, fill=White, right=0.3cm of 4](iex1){$i_{E\mathcal{X}_1}$};
\node[rectangle, draw, fill, right=0.2cm of iex1](4_1){};
\node[above=0.1cm of 4_1](l4_1){\scriptsize 4.1};
\node[rectangle, draw, fill=White, right=0.2cm of 4_1](iex2){$i_{E\mathcal{X}_2}$};

\begin{pgfonlayer}{background}
   \node [draw, fill=Gray!30,fit=(iex1) (iex2) (l4_1)] (Ex){};
\end{pgfonlayer}
\node[below=0.1cm of Ex](lEx){\scriptsize $E.\mathcal{X}$};

\node[rectangle, draw, fill, right=0.5cm of ifx2, yshift=-0.6cm](5){};
\node[above=0.1cm of 5](l5){\scriptsize 5};

\node[rectangle, draw, dashed, inner sep=0.3cm, fit=(1) (lCx)(lAx) (5)](border2){};

\draw[->, Red, thick](1) -- (icx1);
\draw[->, Red, thick](icx1) -- (1_1);
\draw[->, Red, thick](1_1) -- (icx2);
\draw[->, Red, thick](icx2) -- (3);

\draw[->, Red, thick](2) -- (iax1);
\draw[->, Red, thick](iax1) -- (2_1);
\draw[->, Red, thick](2_1) -- (iax2);
\draw[->, Red, thick](iax2) -- (4);

\draw[->, Red, thick](3) -- (ifx1);
\draw[->, Red, thick](ifx1) -- (3_1);
\draw[->, Red, thick](3_1) -- (ifx2);
\draw[->, Red, thick](ifx2) -- (5);

\draw[->, Red, thick](4) -- (iex1);
\draw[->, Red, thick](iex1) -- (4_1);
\draw[->, Red, thick](4_1) -- (iex2);
\draw[->, Red, thick](iex2) -- (5);
\end{tikzpicture}

\vspace{0.2cm}
(b)
\end{center}
\caption{Example: code and control points}
\label{f:cpt}

\end{figure}
\subsubsection{Control flow graph (CFG).} \label{s:cfg}
A control flow graph $G(V, E, \mathcal{N}, \mathcal{X})$ is a graph with a set of vertices/nodes $V$ and (control flow) edges $E$. $V = V_I \cup V_D$. Here,
\begin{itemize}
\item $\mathcal{N}$ is a unique node, called the entry node of the CFG $g$, referred to as $g.\mathcal{N}$.
\item $\mathcal{X}$ is unique node $\mathcal{X}$, called the exit node of the CFG $g$, referred to as $g.\mathcal{X}$. The exit node has no successors.
\item $V_I$ is the set of instruction nodes, i.e. nodes with an instruction in them with a possible side-effect in the action language.  An instruction node $v_I \in V_I$ has zero or one successor (referred to as $v_I.s$).
\item $V_D$ is the set of decision nodes. Each decision node $v_D \in V_D$ has a boolean expression in the action language called the condition, referred to as $v_D.c$. A decision node has two successors, namely the \emph{then} successor (referred to as $v_D.t$) and the \emph{else} successor (referred to as $v_D.e$).
\end{itemize}

Additionally:
\begin{itemize}
\item A CFG may have a single node, in which case, $\mathcal{N} = \mathcal{X}$.
\item $\mathcal{N} \in V$. $V_\mathcal{X} \in V_I$.  
\end{itemize}

\begin{figure}
\begin{tabular}{c @{} c}
\begin{minipage}{0.5\textwidth}
\begin{lstlisting}
x = 0;
while(x < 10)
  x++;
\end{lstlisting}
\end{minipage}
&
\begin{minipage}{0.5\textwidth}
\begin{center}
\scalebox{0.6}{
\begin{tikzpicture}
\node[rectangle, draw](i1){$x \gets 0$};
\node[circle, fill=Gray!50, inner sep=0.1cm, above left=0.1cm of i1](li1){$i_1$};

\node[circle, draw, fill=Black, minimum size=1mm, above=of i1](st){};

\node[ellipse, draw, below=of i1](d1){$x < 10$};
\node[circle, fill=Gray!50, inner sep=0.1cm, left=0.1cm of d1](ld1){$d_1$};

\node[rectangle, draw, below=of d1](i2){$x \gets x+1$};
\node[circle, fill=Gray!50, inner sep=0.1cm, left=0.1cm of i2](li2){$i_2$};

\node[rectangle, draw, line width=1mm, below=of i2](i3){skip};
\node[circle, fill=Gray!50, inner sep=0.1cm, left=0.1cm of i3](li3){$i_3$};

\draw[->, Red, thick](st) -- (i1);
\draw[->, Red, thick](i1) -- (d1);
\draw[->, Red, thick](d1) -- (i2);
\draw[->, Red, thick](i2) -- + (-90:1cm) -- + (210:2cm) |- (i1);
\draw[->, Red, thick](d1) -- +(0:2cm) |- (i3);
\end{tikzpicture}
}
\end{center}
\end{minipage} \\

\multicolumn{2}{c}{
$g.\mathcal{N}=i_1$, $g.\mathcal{X}=i_3$, $g.V_I=\{i_1, i_3\}$, $g.V_d=\{d_1\}$, $i_1.s=d_1$, $d_1.t=i_2$, $d_1.e=i_3$
}
\end{tabular}
\caption{Control flow graph}
\label{f:cfg}
\end{figure}
\begin{example}[Control flow graph] \label{ex:cfg}
\end{example}
\subsubsection{Control Points.} \label{s:cpt}

To compute the sequence in which the code blocks get executed, we can think of a partial order (or the corresponding directed acyclic graph) of code blocks such that code blocks in the same sequence are in that order. The nodes of the code sequence DAG correspond to the leaf nodes of the code containment tree. The ordering relations between various nodes of this DAG can be directly derived from the structure of the code containment tree.

\begin{example}[Code partial order]
The model in Figure~\ref{f:conf-code-example}(a) in configuration $\mathbb{C}= \{A, C\}$ with input event $e$ causes $t_{GN}$ to fire. The code partial order corresponding to this simulation step is shown in Figure~\ref{f:cpt}(a). As can be seen, the nodes of this DAG correspond to the leaves of the code containment tree shown in Figure~\ref{f:cct}(b).

\qed
\end{example}

As we have pointed out, instructions in concurrently composed pieces of code can interleave. Hence, we work with the idea of \emph{control points} -- points in the executing code which are visible to the simulator and are subject to interleaving in case of concurrent composition.

\begin{example}[Control Points]
A portion of the code partial order shown in Figure~\ref{f:cpt}(a) is shown in Figure~\ref{f:cpt}(b). The control points here are: 1, 1.1, 3, 3.1, 2, 2.1, 4, 4.1, 5 and so on. Here, control points like 1, 2, 3, 4, 5 etc. are control points outside individual control flow graphs. However, control points like 1.1, 2.1, 3.1, 4.1 etc. are control points inside individual control flow graphs.

In Figure~\ref{f:cpt}(b), the two sequences -- 1, 1.1, 3, 3.1 and 2, 2.1, 4, 4.1 -- are concurrently composed and execute in two concurrent threads. As is clear from the figure, 1.1 is reached strictly after reaching 1; 3 after 1.1 etc. However, Any of 1.1 or 2.1 may be reached earlier depending on which of the two concurrent threads progresses first. At a point when the execution progresses beyond 3.1 and 4.1, the two threads join and give place to a single thread which is at control point 5.
\qed
\end{example}

During code execution, there can be multiple active threads. The simulator keeps track of its progress through each running thread by maintaining a set of control points $CP$. Every code simulation step will cause the execution of one of the instructions immediately after one of the control points $cp \in CP$. Choice of $cp$ is done randomly. After this, $cp$ gets removed from $CP$ and gets replaced by zero, one or more control points, as the case may be, which succeed $cp$ in the code partial order.

\begin{scriptsize}
\begin{align*}
parent(n) &= n' \text{ \textbf{if} }  n' =  \text{Seq}([... n...]) \text{ \textbf{or} } \text{Conc}(\{... n...\}) \\
&\hspace{0.5cm} nil \text{\textbf{ otherwise.} }\\
first(n) & =\text{ \textbf{ if } } n\text{ \textbf{is a} } CFGCode \text{ \textbf{then} 	} \{n.\mathcal{N}\} \\
& \hspace{0.5cm}\text{ \textbf{ if} } n = \text{Seq}([c_1, c_2, ... c_n]) \text{ \textbf{then} } first(c_1) \\
& \hspace{0.5cm}\text{ \textbf{ if} } n = Conc(\{c_1, c_2, ... ,c_n\})  \text{ \textbf{then} } \bigcup\limits_{c_i \in \{c_1, c_2, ..., c_n\}} first(c_i) \\
nextCFGCodes(c) &= \text{ \textbf{if} } parent(n) = nil \text{ \textbf{then} } \{\}\\
& \hspace{0.5cm}\text{\textbf{if} } parent(n) = Seq([c_1, c_2, ... c_n]) \land c = c_i \text{ \textbf{then} } \\
& \hspace{1cm}\text{ \textbf{if} } c_i \neq c_n \text{ \textbf{then} } first(c_{i+1}) \\ 
& \hspace{1cm}\text{ \textbf{else} } nextCFGCodes(parent(n)) \\
& \hspace{0.5cm} \text{ \textbf{if} } n.parent = Conc(\{c_1, c_2, ... ,c_n\}) \text{ \textbf{then}} \\
& \hspace{1cm} nextCFGCodes(parent(n)) \\
nextCP(n) &=  \text{ \textbf{if} } (n = n.cfg.\mathcal{X}) \textbf{ then } nextCFGCodes(n)\\
& \hspace{0.5cm} {\textbf{else } } succ(n, n.cfg) 
\end{align*}
\end{scriptsize}

The function $nextCP$ computes the set of control points that replace the currently processed control point. An internal control point $cp$ will be replaced by one of its successors in $cp.cfg$, the CFG it belongs to. However, if $cp$ is an exit node of $cp.cfg$, then the next of control points to replace it in $CP$ are the entry node other CFGs. Such CFGs (and their unique entry node) can be identified using the functions $nextCFGCodes$ and $first$. Please note that these are computed through an implicit traversal of code containment tree, discussed in Section~\ref{s:cct}.

When a member $cp'\in nextCP(cp)$ happens to be a join point, (i.e. having multiple predecessor control points), the situation needs to be dealt with separately. This is a point where multiple threads will collapse into one at an appropriate time. This is when all predecessors of $cp'$ have been processed, thus bringing the control points of all these thread to $cp'$. This concludes their execution. Hence, when the first of these threads reaches $cp'$, we start keeping track of the number of threads waiting to join at $cp'$, and $cp'$ is kept waiting. When, all these threads reach $cp'$, we insert $cp'$ into $CP$ scheduling it for execution in the next code simulation step.

Note that due to the complex hierarchy in which the code blocks in a statechart may be arranged at runtime, it may appear that keeping track of them would be very complex. However, with the above approach, we are able to essentially collapse the control front (the control points of all the active threads) into a flat set.

\begin{example}[Code Simulation Trace] \label{ex:cst}
Consider the code partial order shown in Figure~\ref{f:cpt}(b). We track the successive values of the current control point set $CP$ and the join points $JP$ in a typical code simulation.
\begin{align*}
(CP, JP) &= (\{\}, \{\}), (\{1, 2\}, \{\}), (\{1, 2.1\}, \{\}), (\{1, 4\}, \{\}), (\{1.1, 4\}, \{\}) \\
         & (\{3, 4\}, \{\}), (\{3.1, 4\}, \{\}), (\{4\}, \{5 \mapsto \{4.1\}\}), \\
         & (\{4.1\}, \{5 \mapsto \{4.1\}\}), (\{5\}, \{\}), ... (\{7.1\}, \{\}), (\{8\}, \{\}), \\
         & (\{8.1, 8.2\}, \{\}), ... (\{9.1, 10.1\}, \{\}), (\{9.1\}, \{\}), (\{\}, \{\})
\end{align*}

Following points in the above trace are noteworthy:
\begin{itemize}
\item The set $CP$ and map $JP$, both start off empty and end up empty over a simulation step.
\item The join point $5$ is reached when the control front passes control point $3.1$. At this point, an entry is added to $JP$ with $5$ as key and $\{4.1\}$ as value, as $4.1$ is the previous control point to $5$ in the currently active threads (only one in this case, as the thread corresponding to $3.1$ has already finished execution) which will eventually join at $5$.
\item As the control front passes $4.1$, it gets removed from the values of key $5$ in $JP$. This makes the value set of $5$ empty in $JP$. Therefore, $5$ is removed from $JP$, and added to $CP$.
\item Control point $8$ is fork point. $8.1$ and $8.2$ are the internal control points of $L.\mathcal{N}$ and $M.\mathcal{N}$ respectively. As soon as the control front moves past $8$, it is replaced by $8.1$ and $8.2$  in $CP$.
\item Code simulation for this simulation step concludes when $CP$ becomes empty. 
\end{itemize}
\end{example}

\subsubsection{Operational Semantics -- Code Simulation.}
During the code simulation of a simulation step, the code data structure  $\mathtt{C}$ that gets executed remains constant. In this part of the discussion, we will omit mentioning it in most places. The first step of code simulation involves populating the control front $CP$ with appropriate set of CFG Nodes in the code. These are nothing but the initial entry nodes of the initial CFGCodes of the code. In Figure~\ref{f:os-codesim}, we show this in rule \textsc{code-sim-init}. Rule \textsc{code-sim-internal} presents the operational semantics when the currently processed CFG node $n$ is not an exit node of its CFG. Its successor node in the CFG $n'$ is computed as $nextCFGNode(n, \sigma)$. This is a direct successor if $n$ corresponds to an instruction node. However, if $n$ is a decision node, $n'$ depends on valuation of the condition expression $n.c$ of $n$. If $n.c$ evaluates to true in the current value environment $\sigma$, then $n'$ is $n$'s then-successor; otherwise, it is the else-successor of $n$.

The rule \textsc{code-sim-exit-node} handles the case when the current CFG node $n$ chosen randomly from the control front $CP$ (using the function $any$) is the exit node of its CFG. As discussed above, this step could lead to joining (i.e. two threads collapsing into one) or forking (one thread giving place to multiple threads) of code execution. This could also lead us to the completion of code simulation when $CP$ becomes empty again. In this step, all successors of $n$ which are not join points (shown by set $N$), directly get added to $CP$. The other set of successors are join points, shown by set $J = J_i \cup J_c \cup J_o$. $J_i$ is the set of new join points, i.e. for each member of $J_i$, $n$ is the first of the predecessors to have got processed. In this step, all members of $J_i$ get added to $JP$. $J_c$ is the set of those join point successors, one of whose predecessors have already been processed in an earlier code simulation step, and hence, they had previously been added to $JP$. Even after the processing of $n$, there are other predecessors of the members of $J_c$ which await processing; hence all members of $J_c$ continue to be part of $JP$. $J_o$ are those join point successors, all of whose predecessors have completed their processing in this step, $n$ being the last of them. Hence, all members of $J_o$ are removed from $JC$ and inserted to the control front $CP$.

Note that \textsc{code-sim-exit-node} subsumes the case when the control front $CP$ empties out, which means that the code execution halts.

\begin{figure}
\begin{center}
\begin{mathpar}
\inferrule*[lab={\color{Maroon}code-sim-init}, width=\textwidth,
leftskip=2em,rightskip=2em]
{
}
{
\sigma, \{\}, \{\} \longrightarrow \sigma', first(\mathtt{C}), \{\}
}
\end{mathpar}
\begin{mathpar}
\inferrule*[lab={\color{Maroon}code-sim-internal}, width=\textwidth,
leftskip=2em,rightskip=2em]
{
n = any(CP) \\ n' = nextCFGNode(n, \sigma) \\
\sigma \vdash n.inst \Downarrow \sigma' \\
CP' = CP \setminus \{n\} \cup \{n'\}
}
{
\sigma, CP, \{\} \longrightarrow \sigma', CP', \{\}
}
\end{mathpar}
\begin{mathpar}
\inferrule*[lab={\color{Maroon}code-sim-exit-node}, width=\textwidth,
leftskip=2em,rightskip=2em]
{
n = any(CP) \\ n = n.cfg.\mathcal{N} \\
\sigma \vdash n.inst \Downarrow \sigma' \\
next(n) = N \cup J_i \cup J_c \cup J_o \\
N = \{n' | pred(n') = \{n\} \land n' \in CP'\} \\
J_i = \{j |\ |pred(j)| > 1, j \not\in JP \land JP'(j) = pred(j) \setminus \{n\} \} \\
J_c = \{j\ \vert\ |pred(j)| > 1 \land |JP(j)| > 1 \land JP'(j) = pred(j) \setminus \{n\} \}\\
J_o =  \{j |\ JP(j) = \{n\} \} \\
CP' = CP \setminus \{n\} \cup N \cup J_o \\ JP' = JP \cup J_i \setminus J_o
}
{
\sigma, CP, JP \longrightarrow \sigma', CP', JP'
}
\end{mathpar}
\caption{Operational semantics of code simulation}
\label{f:os-codesim}
\end{center}
\end{figure}

\begin{example}[Operational semantics -- Code simulation] \label{ex:oscs}
In Table~\ref{t:oscs}, we present an example of the trace shown in Example~\ref{ex:cst} to illustrate the role of sets $J_i$, $J_c$ and $J_o$.

\begin{table}
\begin{center}
\begin{tabular}{|c|c|c|c|c|c|c|c|c|}
\hline
$CP$ & $JP$ & $n$ & $N$ & $J_i$ & $J_c$ & $J_o$ & $CP'$ & $JP'$ \\
\hline
\hline
$\{\}$ & $\{\}$ & - & $\{\}$ & $\{\}$ & $\{\}$ & $\{\}$ & $\{\}$ & $\{\}$ \\
\hline
$\{1,2\}$ & $\{\}$ & 2 & $\{2.1\}$ & $\{\}$ & $\{\}$ & $\{\}$ & $\{1, 2.1\}$ & $\{\}$ \\
\hline
$\{1, 2.1\}$ & $\{\}$ & 2.1 & $\{4\}$ & $\{\}$ & $\{\}$ & $\{\}$ & $\{1, 4\}$ & $\{\}$ \\
\hline
$\{1, 4\}$ & $\{\}$ & 1 & $\{1.1\}$ & $\{\}$ & $\{\}$ & $\{\}$ & $\{1.1, 4\}$ & $\{\}$ \\
\hline
$\{1.1, 4\}$ & $\{\}$ & 1.1 & $\{3\}$ & $\{\}$ & $\{\}$ & $\{\}$ & $\{3, 4\}$ & $\{\}$ \\
\hline
$\{3, 4\}$ & $\{\}$ & 3 & $\{3.1\}$ & $\{\}$ & $\{\}$ & $\{\}$ & $\{3.1, 4\}$ & $\{\}$ \\
\hline
$\{3.1, 4\}$ & $\{\}$ & 3.1 & $\{\}$ & $\{5\}$ & $\{\}$ & $\{\}$ & $\{4\}$ & $\{5 \mapsto \{4.1\}\}$ \\
\hline
$\{4\}$ & $\{5 \mapsto \{4.1\}\}$ & 4 & $\{4.1\}$ & $\{\}$ & $\{5\}$ & $\{\}$ & $\{4.1\}$ & $\{5 \mapsto \{4.1\}\}$ \\
\hline
$\{4.1\}$ & $\{5 \mapsto \{4.1\}\}$ & 4.1 & $\{\}$ & $\{\}$ & $\{\}$ & $\{5\}$ & $\{5\}$ & $\{\}$ \\
\hline
$\{5\}$ & ... &  & ... & ... & ... & ... & ... & ... \\
\hline

\end{tabular}
\end{center}
\caption{Details of code simulation trace tracking the sets $J_i$, $J_c$ and $J_o$ as per the operational semantics.}
\label{t:oscs}
\end{table}
\qed
\end{example}
\subsection{Computing the New Configuration}
The new configuration $\mathbb{C}'$ is given by the following function:

\begin{align*}
\mathbb{C}' &= \text{\textbf{ if }} \mathcal{T} = \{\} \text{\textbf{ then }} \mathbb{C} \\
&\hspace{0.5cm} \bigcup\limits_{t\in\mathcal{T}} leaves(ST_d(t)) \text{\textbf{ otherwise. }}
\end{align*}
If there are no enabled transitions, the configuration does not change. The event gets lost. If there are one or more enabled transitions, the new configuration is the union of the set of leaf nodes of the destination side state tree of each of these transitions.


\section{Fuzz Testing of Statecharts}
\label{fuzzing}
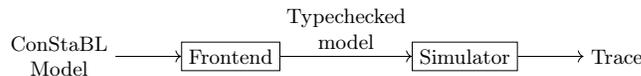
\begin{figure}
\centering
\begin{minipage}{0.8\textwidth}
\resizebox{0.9\textwidth}{!}{
    \centering
	\begin{tikzpicture}
	\node[align=center](cm){ConStaBL \\ Model};
	\node[rectangle, draw, align=center, right=of cm](fe){Frontend};
	\node[rectangle, draw, align=center, right=2cm of fe](sim){Simulator};
	\node[right=of sim](tr){Trace};
	
	\draw[->](cm) -- (fe);
	\draw[->](fe) -- node[align=center, above]{Typechecked \\ model}(sim);
	\draw[->](sim) -- (tr);
	\end{tikzpicture}
}
\end{minipage}   
    \caption{Overview of the ConStaBL Simulator}
    \label{fig:overview}
\end{figure}

We have implemented a simulator as per the operational semantics presented in section \ref{o-semantics}. Fig. \ref{fig:overview} shows an overview of the integral units of the simulator. The grammar for the language \textit{StaBL} has been enhanced to include the \textit{concurrency} construct. The statechart of the system written in ConStaBL and its grammar are given as input to the \textit{parser} that generates the \textit{abstract syntax tree (AST)}. Both the \textit{type checker} and \textit{simulation engine} work on the generated \textit{AST}. The \textit{typechecker} module takes the AST and \textit{structural} semantics as input to indicate any type checking errors. The \textit{simulator} is the simulation engine operate in two modes: \textit{auto} and \textit{interactive}, which implement our \textit{operational} semantics. It works on the \textit{AST}, and the simulation (procedure \textsc{simulate}) consumes a sequence of events, each causing a simulation step (procedure \textsc{simulationStep}). The engine maintains the execution state, configuration, environment, and state at each execution point. Our simulator allows users to simulate the models written in \textit{ConStaBL} and is available in our github repository\cite{2018}.
Statecharts are widely used in various domains\cite{harel2007statecharts}, including safety-critical systems like avionics and automotive control systems. These systems often incorporate subsystems for navigation, obstacle detection, and vehicle communication, resulting in complex interactions and numerous possible execution scenarios. Manual simulation of such systems becomes challenging due to the sheer number of valuations and execution traces. To address this challenge, we have harnessed the advantages of fuzz testing specifically for statecharts. Fuzz testing involves generating a large volume of inputs and observing how the system responds, aiming to uncover software bugs and anomalies. Integrating fuzz testing with statecharts is a novel concept that has not been explored in existing literature. By employing fuzz testing on statecharts, we can leverage several advantages, including:

\begin{enumerate}
\item It aids in evaluating the robustness of a system by subjecting it to a diverse array of inputs, including unusual data, edge cases, and unexpected sequences. This verifies the system's ability to handle such scenarios effectively, without encountering crashes or displaying erroneous behaviour.
\item  It enables rigorous testing and verification by systematically exploring an extensive number of paths in statecharts, which can be practically infinite.  
\item It can be automated, reducing the manual effort required in testing various combinations of interleavings in a concurrent code. It facilitates comprehensive coverage which significantly enhances the detection of challenging concurrency-related bugs that might otherwise remain elusive.
\item The counterexample traces obtained during fuzz testing can be used to easily replicate and analyse the identified issues, further simplifying the debugging process.
\end{enumerate}

Our semantic approach enables a thorough analysis of system models, examining intricate details at the action language level. We detect the following through fuzz testing:
\begin{enumerate}
\item \textbf{Non-determinism} Though we do not utilise priorities, it is still valuable for designers to be aware of the existence of non-determinism (we have explained this in detail in section \ref{o-semantics} - \textit{conflicting transitions}). Detecting non-determinism at compile time based on triggers may result in false-positives, as they depend on the runtime valuation of variables in the $guard$.
\item \textbf{Concurrency conflicts} occur when two enabled transitions within an orthogonal composition attempt to modify the same variable. Detecting such conflicts can be challenging, especially in nested concurrent code where control flow becomes complex. It becomes difficult to manually identify all possible cases in which concurrency conflicts may arise.
\item \textbf{Undesired configurations} can occur in concurrent statecharts when multiple regions enter a combination of states that, while individually valid, lead to undesired behaviour. For example, while it is acceptable for two traffic signals to be independently green, if they are in the same junction displaying opposite signals, it can result in accidents. Therefore, it is crucial to examine the orthogonal composition of traffic lights and determine if there are any scenarios where the configuration becomes $\{green, green\}$.
\item To test the \textbf{reachability} of all states and transitions, which is a common concern in traditional testing. Reachability serves as the counterpart to undesired configuration by aiming to validate the accessibility of all desired states and transitions within the system.
\end{enumerate} 

Fuzz testing, like any testing practice, may not guarantee complete and sound results. However, it is a highly effective technique when applied to statecharts for detecting bugs. Its randomised execution closely resembles the input streams experienced by real-world systems, such as autonomous vehicles. In such systems, multiple subsystems interact with their environment, process inputs from sensors and radars, and control actuators like throttle, steering, and brake. Fuzzing provides a better replication of these scenarios compared to manual simulation. 

Our experiments were conducted using an automotive dataset\cite{juarez2007preliminary} that includes seven subsystems: Cruise Control (CC), Collision Avoidance (CA), Parking Assistant (PA), Lane Guidance (LG), Emergency Vehicle Avoidance (EVA), Parking Space Centering (PSC), and Reversing Assistance (RA). We transformed the dataset from Statemate to Constabl semantics, omitting the transition priorities in our model intentionally to test it in a non-prioritised system. Additionally, we merged the subsystems through orthogonal composition prior to conducting the experiments. While these models initially did not contain any inherent errors, we deliberately injected errors at various locations in the models to evaluate the effectiveness of our fuzzing implementation in detecting them within a short timeframe. 

We integrated Jazzer \cite{jazzer}, a Java-based fuzz testing tool, to perform fuzzing and generate counterexample traces upon encountering failures. We could identify conflicts where two subsystems issue conflicting actions, such as modifying the \textit{Speed} of the vehicle concurrently (same-actuator conflicts). Also, the subsystems can result in undesired configurations like $\{Slow, Mitigate\}$, where $EVA$ instructs the system to slow down where as $CA$ instructs the system to mitigate the risk of colliding by halting the vehicle. As halting the vehicle may hinder the movement of emergency vehicle, this is an undesired configuration. We also injected non-determinism inducing guards in transitions at various levels of hierarchy and could flag them using our simulator at run-time. The experiments were performed on an Ubuntu machine with 16 GB RAM. Simulations were conducted by generating event sets of sizes 5000, 10000, and 20000 using the automotive model consisting of all seven subsystems, which includes approximately 80 states and 175 transitions. The results of these simulations can be found in our github repository\footnote{https://github.com/sujitkc/statechart-verification/tree/skc-simulator-test/src/dfa/outputs/uwfms}.

In addition, we applied our tool to several other examples to assess the reachability from one random configuration to another random configuration using fuzzing as part of our tool's coverage testing. This experimentation has provided us with confidence in the effectiveness of fuzz testing in statechart scenarios. It has also opened up possibilities for employing fuzzing to verify different properties, which could be an area for future research.

Finally, the entire experiments stands testimony to the claim that upstream modelling can allow early validation and detection of bugs. In all these experiments, the models were developed using the semantics presented in this paper proving its practicality. The experiments were run using our prototype simulator proving it a useful engine to implement early stage model validation steps.

\section{Conclusion and Future work}
\label{conclusion}
Statecharts are valuable tools for analysing, designing, and implementing complex systems due to their ability to represent different levels of abstraction and intricate system interactions. In this paper, we introduced ConStaBL, a variant of concurrent statecharts. Our contribution was the development of a semantics and simulator that allow for the interleaved execution of action code, enabling the detection of concurrency-related issues at an early stage. With the emergence of parallel processing systems and distributed execution of entities, there is a need to further enhance our approach to handle parallelism and incorporate analysis methodologies during the design phase. 

We presented a novel way to do fuzz testing on statechart models directly. To the best of our knowledge, this idea has not been tried earlier. This illustrates how to do early detection of defects at an early stage of SDLC using testing. It also demonstrates the applicability of the semantics presented in this paper to model realistic systems, and the ability of our simulator as a powerful aid in analysis.

We aim to leverage the fine-grained nature of our simulator to detect a wider range of defects in systems involving interleaved concurrency. Additionally, we plan to explore extensions discussed in the work \cite{juarez2007preliminary}, such as establishing an accepted threshold for actuator conflicts by comparing variable valuations in the value environment. These extensions will be a focal point of our future studies.

\bibliographystyle{splncs04}

\bibliography{refs}

\appendix
\section{Glossary of Terms}
\begin{itemize}
\item $\mathcal{M}$: The statechart model
\item $S$: The set of states
\item $s_1, s_2, ...$: states
\item $C$: Configuration
\item $CST(C)$: Configuration state tree for configuration $C$
\item $\sigma$: Value environment
\item $\mathbb{T}_{tr}(n)$: Subtree of node $n$ in tree $tr$ 
\item $\widehat{\mathbb{T}}_{tr}(n, C)$: Subtree of node $n$ in tree $tr$ sliced with $C$, a subset of all leaf nodes of $tr$
\item $ST_s(t, C)$: Source side state tree for enabled transition $t$ in configuration $C$ 
\item $ST_d(t, C)$: Destination state tree for enabled transition $t$ in configuration $C$
\item $CT_s(t, C)$: Source side code tree for enabled transition $t$ in configuration $C$ 
\item $CT_d(t, C)$: Destination side code tree for enabled transition $t$ in configuration $C$ 
\item $code_s(t, C)$: Source side code for enabled transition $t$ in configuration $C$ 
\item $code_d(t, C)$: Destination side code for enabled transition $t$ in configuration $C$
\item $code(t, C)=code_s(t, C)\ t.a\ code_d(t, C)$: Code for enabled transition $t$ in configuration $C$ 
\item $\mathcal{C}(t, C)$: The code that gets executed when an transition $t$ in configuration $C$ is fired.  
\item $\mathcal{C}_s(t, C)$: The source side code that gets executed when an transition $t$ in configuration $C$ is fired.
\item $\mathcal{C}_d(t, C)$: The source side code that gets executed when an transition $t$ in configuration $C$ is fired.
\end{itemize}

\end{document}